\documentclass[12pt]{article}
\usepackage[russian,english]{babel}
\usepackage{amssymb,graphicx,wrapfig}
\author{Vladimir A. Petrov\footnote{e-mail: Vladimir.Petrov@ihep.ru}
and Nikolai P. Tkachenko\footnote{Nikolai.Tkachenko@ihep.ru}}
\index{}
\title{Coulomb-Nuclear Interference: Theory and
Practice for $pp$-Scattering at 13 TeV.}
\date{}

\begin{document}

\maketitle
\begin{center}

A.A. Logunov Institute for High Energy Physics 

NRC "Kurchatov Institute", Protvino, RF
\end{center}
\begin{abstract}
We provide a detailed reconsideration of the  theoretical basis for the treatment
of Coulomb-nuclear interference (CNI) and a corresponding thorough analysis
of the procedure of extraction of the basic parameters
$\rho^{~pp}, \sigma_{tot}^{~pp}$ and $B^{~pp}$ from the TOTEM data at 
$\sqrt{s} = 13$ TeV. A more substantiated account of CNI, as well as an
in-depth statistical analysis of the TOTEM data at low transferred
momenta, give results that differ from those published by the TOTEM
collaboration.
\end{abstract}

\section*{Introduction}

Measurements by the LHC TOTEM Collaboration at 13 TeV \cite{TOT} caused
a number of papers with vivid discussions, often with opposite
conclusions. The TOTEM Collaboration gives two values\footnote{As said in \cite{TOT} "depending on different physics assumptions and mathematical modelling". We comment on the latter reason in Section 3 and  Conclusions. }
of the parameter
$$
\rho = \frac{\mbox{Re} T_{N}(s,0)}{\mbox{Im} T_{N}(s,0)} =
0.09\div 0.10
$$ 
with a strikingly small error\footnote{Here $T_{N}(s,t)$
stands for the $pp$ ("nuclear") scattering amplitude
in absence of electromagnetism. As usual, $s$ stands for the c.m.s. energy
squared, white $t$ is the invariant momentum transfer squared.}, less than
$10\%$, although the very data show large systematic errors.
An unexpectedly low value of $\rho$ given by the TOTEM  prompted some authors
to consider these results as a proof of the existence of  the so-called "Maximal
Odderon" (in particular those of Ref.\cite{Nic}, in which $\rho$ was estimated
as very close to $ 0.1 $). Such an interpretation was adopted in \cite{TOT}
(with some reservation, though) as a first experimental observation of a
"3-gluon compound state".

It was also claimed in \cite{TOT} that such a low value of $ \rho $ along
with the mentioned narrow error corridor supposedly "has implied the exclusion
of all the models classified and published by"the COMPETE Group \cite{PRED}.

 Concerning the credibility of the statement about the Odderon observation
we refer the reader to Refs. \cite{Ptr} and \cite{Ptrr}where a
detailed discussion of this problem is given. 
 
In the present work we concentrate on two important issues of the parameter
retrieval from the data.
\begin{itemize}
\item First of all, this is a way of taking into account CNI effects in
the full scattering amplitude. Beyond a simplistic adding the Coulomb 1-photon
exchange to the nuclear amplitude (which is obviously theoretically untenable)
we find in the literature two ways of accounting for CNI. 

The first one goes back to the pioneering work of H.Bethe and was significantly
improved by D.R. Yennie and G.B. West \cite{Be}. The idea is that the inclusion
of Coulomb exchanges leads to the appearance of an additional phase in the
strong interaction scattering amplitude. Within  a couple of decades this
method served a convenient phenomenological tool for the description of the
differential cross-section with account for CNI and extracting the parameters
mentioned above (see e.g.\cite{PPP}).

Nonetheless, afterwards it was criticized by R.Cahn (and a dozen years later
in a more detailed way by V. Kundr\'{a}t and M. Lokaj\'{i}\v{c}ek) \cite{Cahn}.
It was noted that the Bethe-Yennie-West method requires the phase of the
nuclear scattering amplitude $ T_{N} $ to be independent of the momentum
transferred which is a rather restrictive condition. Moreover, as was proved
recently \cite{PePh} , such an independence, even in an arbitrary narrow
interval, leads to identical vanishing of this amplitude.In addition, it
is noted in \cite{Don} that the form of the Coulomb phase proposed by Yennie
and West contradicts the general properties of analyticity in the t-channel.

As a replacement for the Bethe-Yennie-West method, a different method was
proposed in \cite{Cahn} for taking into account the CNI effects, which does
not impose unnecessary restrictions on the phase of the nuclear scattering
amplitude. Exactly this method was adopted  in \cite{TOT} for retrieving
the $\rho$-parameter etc which attracted a close attention.

However, this last method, which we will call CKL
(for Cahn-Kundr\'{a}t-Lokaj\'{i}\v{c}ek), although it has been qualified in \cite{TOT} 
as " the most general interference formula", also turned out to be flawed. 
The problem was, in particular, in an oversimplified way of the form
factor account.

A formula devoid of this shortcoming was derived in Refs. \cite{PTR}, \cite{PSt}.
  
In what follows we use an expression based on these latter papers for a correct
account for CNI effects. We believe it is suitable for general use.
\item The aforementioned low value of $ \rho $ was obtained in \cite{TOT}
on the basis of an analysis  of the differential cross-section at low values
of the transferred momenta. To this end two arrays of the data were chosen.
Specifically, the sets with $|t| \leq 0.15~\mbox{GeV}^{2}$  and with
$|t| \leq 0.07~\mbox{GeV}^{2}$, though no reason why namely these values
of cut-off were taken was given.

Following the method of  Ref.\cite{TOT},  we also use, when processing the
data, a cut-off in $|t|$ but find that together with an upper cut-off one
has to impose a lower cut-off as well, and in Section 3 we motivate our choice
for both. The values of parameters ($\rho$, total cross-section
$\sigma_{\mbox{tot}}$ and the forward slope $B$ of $d\sigma/dt$), retrieved
with help of such a procedure, turn out to be different from those announced
in \cite{TOT}.As regards the $\rho $ parameter,  we find that the statistically
justified retrieval leads inevitably to a significant widening of the
error corridor what is of fundamental importance.
\item The last item we would like to draw the attention to is the question:
to which extent the smearing of the proton electric charge over its volume
(embodied by the form factor $\mathcal{F}$)  makes the parameters $\rho$
etc different from the idealized case of the point-like charge
($\mathcal{F}=1$)? The answer turned out to be quite surprising and quite
helpful for  estimate of accuracy of the cross-section description in the
realistic case.
\end{itemize}
\newpage
\section{ Exact formula}

If we assume, as it is generally done, that the elastic S-matrix
$S_{C+N}(b)$ in impact parameter space factorizes in strong and Coulombic
interactions\footnote{Actually this property cannot be
exact and rather holds at large impact parameters, relevant to the
subject of the present discussion.}
$$
~~~~~~~S_{C+N}(b) = S_{C}(b)S_{N}(b)
$$
 then we can obtain without much difficulty the following
formula for the modulus of the full scattering amplitude $ T_{C+N} $(for the sake of brevity we will mostly omit the explicit
indication of the "passive" parameter $s$ as an argument)
\begin{equation}
\mid T_{C+N}\mid_{q\neq0}^{2} =
4s^{2} S^{C} (q,q)+
\int\frac{d^{2}q'}{(2\pi)^{2}}\frac{d^{2}q''}{(2\pi)^{2}} S^{C} (q',q'')T_{N} (q-q')T_{N}^{\ast} (q-q'')+~~~~~~~~~~~~~~~~~~~~~~
\label{eq1}
\end{equation}
$$
~~~~~~~~~~~~~~~~~~~~~~~~~~~~~~~~~~~~~~~~~~~~~~~~~~~~~~~~~~
+4s \int\frac{d^{2}q'}{(2\pi)^{2}} \mbox{Im}[S^{C} (q,q')T_{N}^{\ast} (q-q^{'})]
$$
where
$$S^{C} (q^{'},q^{''})= \int d^{2}b^{'}d^{2}b^{''}
e^{i{q}^{'}{b}^{'}-i{q}^{''}{b}^{''}} e^{2i\alpha
\Delta_{C} ( b^{'},\, b^{''})}
$$
and
$$\Delta_{C} ( b^{'},\, b^{''}) = \frac{1}{2\pi}\int d^{2}k
\frac{\mathcal{F}^{2}(k^{2})}{k^{2}}(e^{-ib^{''}k} - e^{-ib^{'}k})=
\int_{0}^{\infty}\frac{dk}{k}\mathcal{F}^{2}(k^{2})[J_{0}(b^{''}k)-J_{0}(b^{'}k)].
$$
The fact that we are dealing directly with the square of the amplitude modulus,
and not with the amplitude itself, allows us to avoid the well-known IR divergence
that resides in the phase $\mbox{Arg} T_{C+N}$, and, along with this, dangerous
manipulations with expressions that diverge when removing the IR regularization.
Equation (\ref{eq1}) and the expression for $\Delta_{C}(b^{'}, b^{''})$ contain
only convergent integrals both in IR and in UV. 
In this paper we mostly use, instead of $t$, a more convenient variable
$$
\textbf{q}^{2} \equiv q^{2} \equiv q^{2}_{\perp} = ut/4p^{2} = p^{2}\sin^{2}\theta,
\; s = 4p^{2}+ 4m^{2},
$$
which reflects the $ t-u $ symmetry of the $ pp $ scattering.
At $\theta\rightarrow 0\;\quad q^{2} \approx -t$ while at
$\theta\rightarrow \pi\;\;\quad q^{2} \approx -u$.
We will use the same notation $q$ both for 2-dimensional vectors
$\textbf{q}$ and their modules $\vert \textbf{q} \vert$.
In the latter case, the limits of integration are indicated explicitly.

In Eq.(\ref{eq1}) we impose the condition $ q\neq0 $ which corresponds to
real experimental conditions (the scattered proton cannot be detected arbitrarily
close to the beam axis). The "forward" observables are understood as a result
of extrapolation $t\rightarrow 0$:
\begin{equation}
\rho = \lim_{t\rightarrow0}  \frac{\mbox{Re} T_{N}(s,t)}{\mbox{Im}T_{N} (s,t)},
~~~
\sigma_{\mbox{tot}}\cdot(\hbar c)^{-2} =
\lim_{t\rightarrow0}\frac{\mbox{Im} T_{N}(s,t)}{2p\sqrt{s}},
~~~
B =\lim_{t\rightarrow0}\frac{d[ \ln(d \sigma_{N}/dt)]}{dt} (s,t),
\label{eq2}
\end{equation}
where
$$
\frac{d\sigma_{N}}{dt} \left[\frac{\mbox{mb}}{\mbox{GeV}^{2}}\right]=
\frac{(\hbar c)^{2}}{16\pi s^{2}} \mid T_{N}\mid^{2}, \:
(\hbar c)^{2} = 0.389379...\mbox{ [mb}\cdot \mbox{GeV}^{2}].
$$

However, this does not concern expressions appearing as integrands when
they may very well contain terms like $\delta (\textbf{q})$ with a support
of measure zero.

As we deal with high energies and have in the integrands fast decreasing
nuclear amplitudes $ T_{N}(k^{2}) $ and form factors $\mathcal{F}(k^{2})$
we can (modulo vanishingly small corrections) extend the integration in
$|\textbf{k}|$ (kinematically limited by 
$\sqrt{s}/2$) over the whole 2D transverse momentum space. The benefit is
a convenient opportunity  to freely use direct and reversed 2D Fourier transforms.

Note that the "Coulomb kernel" $S^{C} (q^{'},q^{''})$ has simple boundary
properties
$$
S^{C} (q^{'},q^{''})\mid_{\alpha=0} =
(2\pi)^{2} \delta (\textbf{q}^{'})(2\pi)^{2} \delta (\textbf{q}^{''})
\mbox{~~~while~~~}
\int S^{C} (q^{'},q^{''}) d^{2}q^{'}d^{2}q^{''}/(2\pi)^{4}  = 1, \;
\forall \alpha .
$$
In principle, when applying to the data analysis, one could deal
directly with Eq.(\ref{eq1}) which is an all-order (in $\alpha$) exact expression
free of singularities.

However, in general, it is hardly possible to obtain an explicit and "user
friendly" analytic expressions with arbitrary $T_{N}$ and $\mathcal{F}$ which would
allow their convenient practical use. 

So, in practice we have to deal rather with expansions in the fine structure
constant $\alpha$. As $\alpha^{2}\approx 5.3\cdot 10^{-5}$ and
$\alpha^{3}\approx 3.9 \cdot 10^{-7}$ it seems that we can fairly limit our
considerations with terms up to $\alpha^{2}$ inclusively with possible uncertainty
not exceeding 1 percent in the worst case\footnote{See Appendix C for more
details.}.

\section{Differential cross-section in
$\mathcal{O}(\alpha^{2})$ approximation}

As was said above\footnote{In this paper we use the symbol
$\mathcal{O}(\alpha^{n})$ to designate (probably with a variance from the
standard mathematical use) the sum of the first $n$ terms of the
Maclaurent expansion in $\alpha$.}, with a non trivial form factor the
all-order expression for the Coulomb kernel $S^{C} (q^{'},q^{''})$ is very
complicated and practically useless\footnote{This is not the case for the
idealized electrically point like protons with $\mathcal{F} = 1$,
see Appendix B.}. Nonetheless, perturbative expansion in $\alpha$, if
to retain at least terms up to $\alpha^{2}$ inclusively, gives a very
precise estimate of $S^{C} (q^{'},q^{''})$ and hence of the differential
cross-section.

Thus we will use the following expansion
$$
S^{C}(q',q'') = (2\pi)^{2} \delta (\textbf{q}^{'})(2\pi)^{2}
\delta (\textbf{q}^{''})+
2i\alpha (2\pi)^{3}[\frac{\mathcal{F}^{2}(q'')}{q''^{2}}
\delta (\textbf{q}^{'})- \frac{\mathcal{F}^{2}(q')}{q'^{2}}
\delta (\textbf{q}^{''})]-~~~~~~~~~~~~~~~~~~~~~~~~~~~~~~~
$$
$$
- 2\alpha^{2}(2\pi)^{2}\int\frac{d^{2}k}{k^{2}}
\mathcal{F}^{2}(k)\frac{d^{2}p}{p^{2}}
\mathcal{F}^{2}(p)[\delta(\textbf{q}^{'})\delta(\textbf{q}^{''}-
\textbf{p} -\textbf{k})-\delta(\textbf{q}^{'}-\textbf{k})
\delta(\textbf{q}^{''}-\textbf{p})-
$$
\begin{equation}
~~~~~~~~~~~~~~~~~~~~~~~~~~~~~~~~~~~~~~~~~~~~~~~~~~~~~~~~~
-\delta(\textbf{q}^{'}-\textbf{p})\delta(\textbf{q}^{''}-
\textbf{k}) +\delta(\textbf{q}^{''})\delta(\textbf{q}^{'}-\textbf{p} -
\textbf{k})]+... 
\label{eq3}
\end{equation}

Expansion (\ref{eq3}) allows us to obtain the following expression for the
differential cross-section
\begin{equation}
(\hbar c)^{-2}\frac{d\sigma_{C+N}}{dt}= \frac{1}{16\pi s^{2}} \mid T_{C+N}\mid^{2}=
J_{0}+ \alpha J_{1} + \alpha^{2} J_{2}
\label{eq4}
\end{equation}
$$
J_{0}= \frac{\mid T_{N}\mid^{2}}{16\pi s^{2}}; ~~~~
J_{1} = -\frac{\mathcal{F}^{2}(q)}{q^{2}}\frac{\mbox{Re} T_{N}(q)}{s} -
\frac{1}{8\pi^{2} s^{2}}\int dk^{2}\frac{\mathcal{F}^{2}(k)}{k^{2}} dp^{2}
( -\lambda (q^{2},k^{2},p^{2}))^{-1/2}_{+}\mbox{Im}(T_{N}(q^{2})T^{*}_{N}(p^{2}))
$$
\begin{equation}
J_{2} = 4\pi \frac{\mathcal{F}^{4}(q^{2})}{q^{4}} + \frac{1}{2\pi s q^{2}}
\int\frac{dk^{2}}{k^{2}}
\frac{dp^{2}}{p^{2}}( -\lambda (q^{2},k^{2},p^{2}))_{+}
^{-1/2}\cdot ~~~~~~~~~~~~~~~~~~~~~~~~~~~~~~~~
~~~~~~~~~~~~~~~~~~~~~~~~~~~~~~~~~~~~~
\label{eq5}
\end{equation}
$$
~~~~~~~~~~~~~~~~~\cdot\left[
q^{2} \mbox{Im}T_{N}(q^{2})\mathcal{F}^{2}(k^{2})\mathcal{F}^{2}(p^{2})-
k^{2} \mbox{Im}T_{N}(k^{2})\mathcal{F}^{2}(p^{2})\mathcal{F}^{2}(q^{2})-
p^{2}\mbox{Im}T_{N}(p^{2})\mathcal{F}^{2}(k^{2})\mathcal{F}^{2}(q^{2})\right]+
$$
$$
~~~~~~~~~~~~~~~~~~~~~~~~~~~~~~~~~~~~~~~~~~~~~~~~~
+\frac{1}{16 s^{2}\pi^{3}}
\mbox{Re}
\left[ \int\frac{dk^{2}\mathcal{F}^{2}(k^{2})dp^{2}
\mathcal{F}^{2}(p^{2})dk'^{2}dp'^{2}}{k^{2}p^{2}}\left\{ ( -
\lambda (q^{2},p^{2},p'^{2}))_{+}
^{-1/2}\cdot\right.\right.
$$
$$
~~~~~~~~~~~~~~~~~~~~~~~~~~~~~~~~~~~~~~
\left.\cdot \left. \left[ (-\lambda (q^{2},k^{2},k'^{2}))_{+}
^{-1/2} T_{N}(k'^{2})T^{*}_{N}(p'^{2})-(-\lambda (k^{2},k'^{2},p'^{2}))_{+}
^{-1/2} T_{N}(q^{2})T^{*}_{N}(k'^{2})\right]\right\}\right] .
$$

The function $( -\lambda (x,y,z))_{+}
^{-1/2} $  with $ \lambda (x,y,z)=x^{2} +y^{2} +z^{2} -2xy -2 xz -2 yz$
is symmetric an all variables and has the properties
$$
~~~~~~~~~~~~~~~~~~\lim_{x=0}( -\lambda (x,y,z))_{+}
^{-1/2}= \pi \delta (y-z)
~~\mbox{and}~~
\int dx( -\lambda (x,y,z))_{+}
^{-1/2} = \pi
$$
which evidently hold cyclically w.r.t. $ x,y,z $.

Let us also remind the definition of the generalized function
$(-\lambda (x,y,z))_{+}^{-1/2}$ \cite{Gel}:
$$
(-\lambda)_{+}^{-1/2} =
\left\{
\begin{array}{ll}
0 & \mbox{for~} \lambda \geq 0 \\
| \lambda |^{-1/2} & \mbox{otherwise} \\
\end{array}
\right.
$$
The properties of $ ( -\lambda)_{+} ^{-1/2} $ help to check the IR convergence
of the integrals in the above formulas.

In what follows we use a phenomenological and often utilized parametrization
of the strong interaction amplitude\footnote{Such an amplitude
has the phase independent on the transferred momentum $q$. As was proved
in Ref. \cite{PePh}, if such a property would exact even in an arbitrary
small but fixed interval of $q^{2}$ it would make the amplitude  $T_{N}(q^{2})$
identical zero. We suppose that Eq.(\ref{eq6}) holds only approximately at
very high energies.}
\begin{equation}
T_{N}(s, q^{2}) = s\sigma_{\mbox{tot}}(s)[ i+ \rho(s)]e^{-B(s)q^{2}/2}.
\label{eq6}
\end{equation}
In terms of this parametrization (which was also essentially  used by the
TOTEM Collaboration in \cite{TOT} for the retrieval of the forward parameters
(\ref{eq2}))  the differential cross-section (\ref{eq4})
acquires the form\footnote{Below, in front of the first
term of Eq. (\ref{eq7}), we omit the factor
$1 +\mathcal{O}(\alpha^{2}) = 1.00004$ assuming it to be just $1$.}  
\begin{equation}
\frac{d\sigma_{C+N}}{dt}=
\frac{\sigma_{\mbox{tot}}^{2}(1+\rho^{2})}{16\pi(\hbar c)^{2} }e^{-B(s)q^{2}}
- \alpha \frac{\rho \sigma_{\mbox{tot}}}{q^{2}}
\mathcal{F}^{2}(q^{2})e^{-B(s)q^{2}/2} +
~~~~~~~~~~~~~~~~~~~~~~~~~~~~~~~~~~~~~~~~~~~~~~~~~~~
~~~~~~~~~~~~~~~~~~~~~~~~~~~~~~~~~~~
\label{eq7}
\end{equation}
$$
\boldmath{~~~~~~~~~~~~~~~~~~~~~~~~~~~~~~~~~~~~~~~~~~~~~~~~~~~~~~~~~~~~~
+ \alpha^{2} \left[ \frac{4\pi (\hbar c)^{2} }{q^{4}}
\mathcal{F}^{4}(q^{2})-\frac{ \sigma_{\mbox{tot}}}{q^{2}}
\mathcal{F}^{2}(q^{2})e^{-B(s)q^{2}/2} H (q^{2})\right]}
$$
Here
\begin{equation}
H (q^{2}) = \ln \left( \frac{\Lambda^{2}}{q^{2}}+1\right) -\sum_{k=1}^{3}\frac{1}
{k \left( 1+\frac{q^{2}}{\Lambda^{2}}\right)^{k}}-\int_{0}^{\infty}
\frac{(1-e^{-\frac{B(s)
\Lambda^{2}}{2}x}I_{0}(B(s)q\Lambda\sqrt{x})}{(1+x)^{4}} dx
\label{eq8}
\end{equation}
In what follows we use the dipole parametrization of the form factor
$\mathcal{F}^{2}(q^{2})= (1+ q^{2}/\Lambda^{2})^{-4}$
with $\Lambda^{2} = 0.71\mbox{ GeV}^{2}$. $I_{0}(z)$ is
the modified Bessel function of the first kind of zero order.

In the absence of an exact expression for $d\sigma_{C+N}/dt$  it
is difficult to judge the accuracy of the approximation as given by
Eq.(\ref{eq7}).

Nonetheless, there is some means for such an evaluation. The matter is that
in the  idealized case of "electrically point like" protons, i.e. if
$\mathcal{F} = 1$, we can obtain ( see Appendix B) an exact expression for
$d\sigma_{C+N}/dt$  which allows us to compare the exact and approximate
(up to $\alpha^{2}$ terms inclusively) expressions. This allows us to judge
how large the approximation error is, at least in the idealized case. Surprisingly,
it appears that the cross-section with a realistic form factor is extremely
close to that in a point like case at low values of $|t|$ we consider.
So it is enough to use the estimation of the accuracy of the second-order
approximation in the point like case as done in Appendix B. The deviation
from the $\mathcal{{O}}(\alpha^{2})$ - approximation due to terms
$\sim \alpha^{3}$ does not exceed a fraction of a percent
(see Appendix C).

In the next Section Eqs.(\ref{eq7}) -- (\ref{eq8}) will be used for retrieving
the parameters $\rho, \sigma_{\mbox{tot}}, B$ from the TOTEM data at
$\sqrt{s}= 13$ TeV.
\newpage
\section{On some problems of the data processing. Retrieval of the parameters
{$\rho^{~pp}, \sigma_{\mbox{tot}}^{~pp} $} and {$ b^{~pp} $} from the TOTEM
measurement of {$d\sigma^{~pp}/dt$}.}

\begin{wrapfigure}[45]{l}{105mm}
\vspace{-6.1mm}\includegraphics[width=105mm]{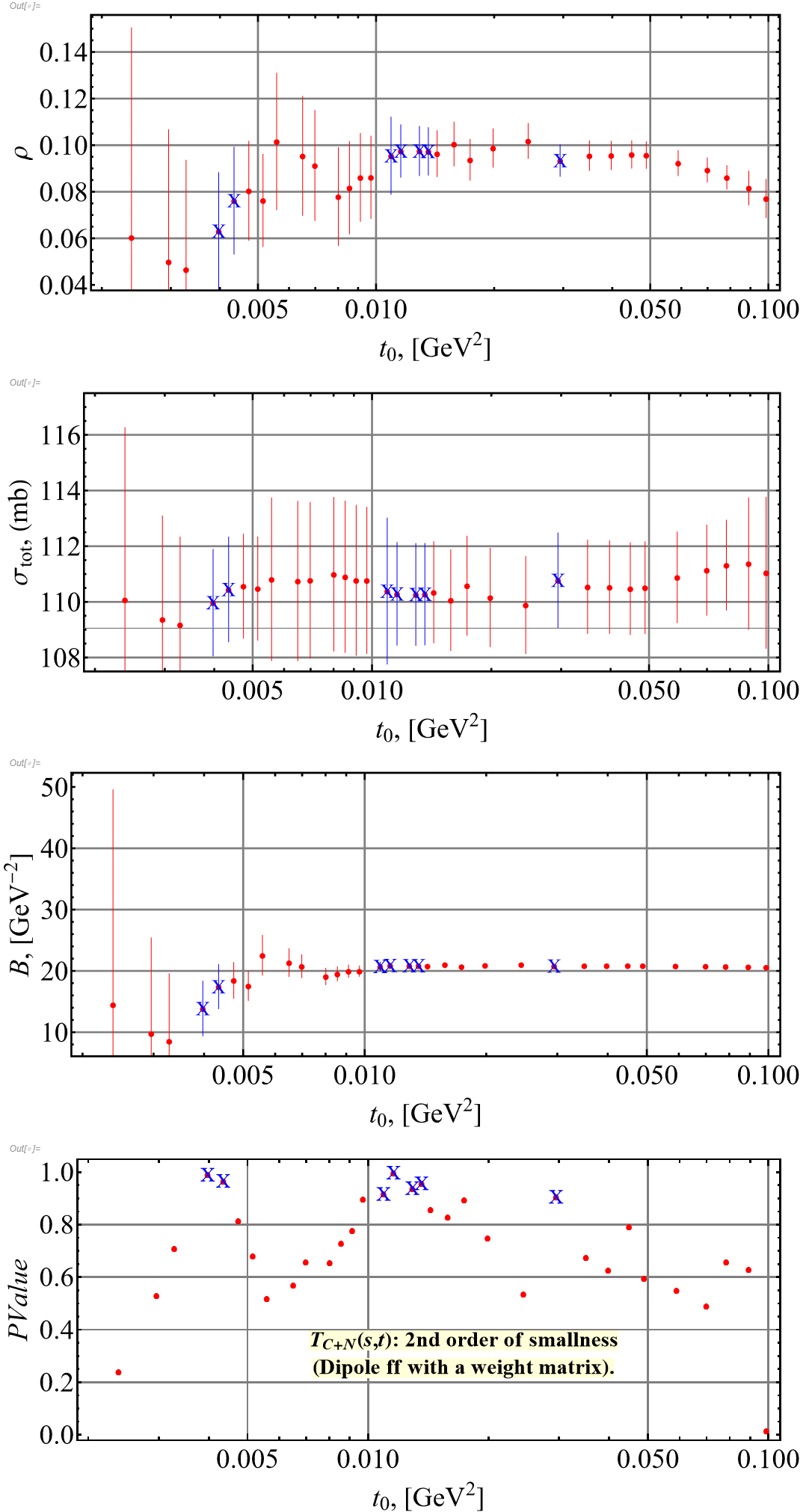}
\vspace{-8.6mm}
\caption{\textit{The values of parameters as functions of the upper cut-off}
$t_{0}$. {\it Points with p-values $> 0.9$ (high level of confidence) are
marked by crosses.}}
\label{Pic1}
\end{wrapfigure}

Before the treatment of the data on the basis of Eqs.(\ref{eq7}),
(\ref{eq8}) we find it appropriate and helpful to highlight and discuss some
important features of the fitting procedures.\vspace{-2.1mm}

\subsection{The $\chi^{2}$-criterion with use of the weight matrix.}

When dealing with the data  \cite{TOT} we compile the weight matrix in
the usual way, as the authors of the TOTEM experiment did. 
Function $ \chi^{2} $ was compiled not for all points. Points with
$\mid t\mid > t_{0}> 0$ were discarded, so that the value of parameters
$\rho, \sigma_{\mbox{tot}},B $ obtained from the  fitting would better
correspond to their definition as "forward observables".

The results are presented in Fig. \ref{Pic1}.

\begin{wrapfigure}[18]{l}{111mm}
\vspace{-6.1mm}
\includegraphics[width=113mm]{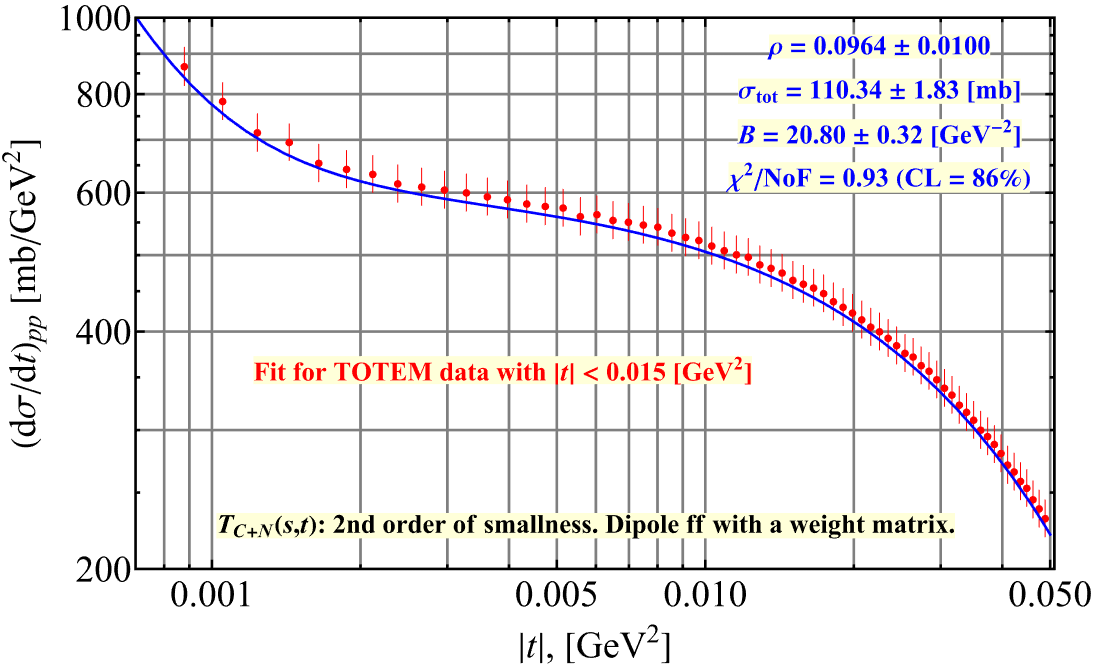}
\vspace{-8.6mm}
\caption{{\it Results of the  fit of the data} \cite{TOT}
{\it at $\sqrt{s} = 13$} TeV {\it at}
$|t| \leq t_{0} \cong 0.015 \mbox{ GeV}^{2}$. {\it Theoretical curve passes
systematically  below the experimental values} ("PPP effect"). {\it Full
errors were obtained using the covariance matrix.}}
\label{Pic2}
\end{wrapfigure}

We have analyzed only the fits with p-values $> 0.9$
(high level of confidence). One can see in Fig. \ref{Pic1} that the points
in the left part of it are unwanted because of their large relative errors.
It is seen that in the interval $|t| > 0.01~\mbox{GeV}^{2}$ a
stabilization of the parameter values occurs and the corresponding values
have practically the same and the smallest errors. This is why we dealt with
$t_{0}\cong 0.015~\mbox{GeV}^{2}$ which corresponds to the procedure
used in our paper \cite{Ezh}. Results of such fitting  with corresponding
parameters  are presented in Fig. \ref{Pic2}.
     
The parameters so retrieved are as follows\vspace{-2.1mm}


$$\rho= 0.10\pm 0.01,$$
$$\sigma_{\mbox{tot}}= 110.3 \pm 1.8~ [\mbox{mb}],$$
$$B = 20.87 \pm 0.35~ [\mbox{GeV}^{-2}].$$

However, as is well seen in Fig. \ref{Pic2},
the fitting curve for the differential cross-section passes systematically
below the average values of experimental points which witnesses that there
is an over(under)estimation of the systematic errors\footnote{Such a systematic
shift was called "Peele's pertinent puzzle" (PPP)(see the paper \cite{ppp}).
So one should take the above values of parameters as evaluative, not as
conclusive ones}.

\newpage

\subsection{The $\chi^{2}$-criterion with use of the "method of the shifting
experimental data" (variant 1).}

Usually, in the absence of correlations, $ \chi^{2} $  is compiled by the
formula:\vspace{-3.1mm}
$$
\chi^{2} = \sum_{i=1}^{N} \left[\frac{Y^{\mbox{th}}(x_{i})-Y^{\mbox{exp}}(x_{i})}
{\sqrt{{\Delta}Y^{2}_{\mbox{stat}}(x_{i})+\Delta Y^{2}_{\mbox{syst}}(x_{i})}}
\right]^{2}\vspace{-1.1mm}
$$
In the case of correlation this method is not suitable and the function
$\chi^{2}$ is compiled using the weight matrix. At the same time, in
our case, the experimental data have systematic errors\footnote{In this case,
the systematic error of all experimental data is very close to $0.055$
of the measured value.}  which significantly exceed the statistical error.
In this variant, the case is often realized when the value
$\chi^{2}/\mbox{NoF}$ turns out to be close to zero, which is
statistically unreliable. 

If the experimental data allow us to build a weight matrix, then  results
are often obtained with the final curve systematically passing below or above
the experimental points, which is realized in our case considered above.
\vspace{-0.1mm}

There is another way to process experimental data when the use of systematic
errors allows a shift of the central values of experimental points. There
can be many ways to shift the experimental data. Let us consider one of them.

We shift the central values by an amount proportional
to the systematic errors and the proportionality coefficient  is
assumed to be the same and equal to $\lambda$ for all points.
$\chi^{2}$ should be written now as:
$$
\chi^{2} = \sum_{i=1}^{\infty} \left[
\frac{Y^{\mbox{th}}(x_{i})-[Y^{\mbox{exp}}(x_{i})+
\lambda \Delta Y_{\mbox{syst}}(x_{i})] }
{{\Delta}Y_{\mbox{stat}}(x_{i})}\right]^{2} + \lambda^{2},
$$
and the parameter  $ \lambda $ is the same for all points of the experimental
data array. In this particular case, $ \lambda^{2} $ is a penalty function
for $ \chi^{2} $ arising in statistics as a consequence of a shift in the
experimental data on systematic errors. I.e. we shift the experimental values
up or down by the amount of their systematic errors with a certain factor
$\mid\lambda\mid \leq 1$, the same for all experimental points.
 In this case we leave in the denominator of the expression for
$\chi^{2}$ only the statistical error of the measured value. The factor
$\lambda$ itself is a fitting parameter.

\begin{wrapfigure}[25]{l}{105mm}
\vspace{-3.6mm}
\noindent\includegraphics[width=106mm]{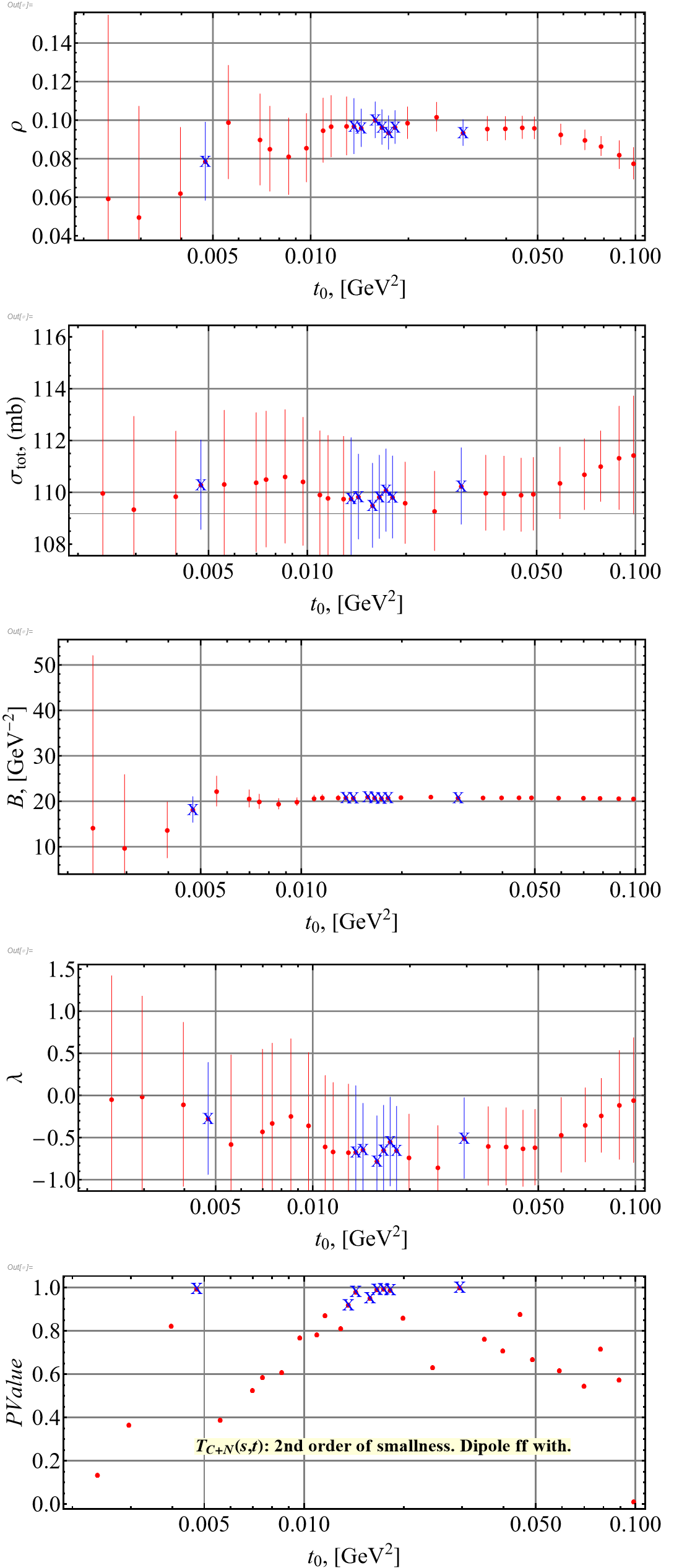}
\vspace{-8.6mm}
\caption{\textit{Parameter values for different cut-off values of experimental
data $t_{0}$. Points with p-value $\geq 0.9  $ are marked with crosses.}}
\label{Pic3}
\end{wrapfigure}

Fitting in this case should not differ significantly from fitting when used
the weight matrix for $ \chi^{2} $ and its results (according to
Eqs. (\ref{eq7}) and (\ref{eq8})) are presented in Fig. \ref{Pic3}.

Here we have (see Fig. \ref{Pic3}) five points with a very high level of
confidence  and the more a point lies to the left, the more it satisfies
the definition of our basic parameters as given by Eqs.(\ref{eq2}). 

As above we use only those fits which have p-values $\geq 0.9$.
The exact values of the parameters with a high level of confidence (blue
dots), taking into account their errors do not allow to give preference to
their values for different values of $ t_0 $. Of course, lower values of
$t_0$ are more preferable, but they have a large error\footnote{At very low
$t_{0}$ the behaviour of all parameters as functions of $t_{0}$ becomes very
unstable.}. For large $t_0$ the errors are
small, but taking them into account reduces compliance with the definitions
of the parameters in question as given in Eqs.(\ref{eq2}).

\newpage 

\begin{figure}[h]
$$\includegraphics[width=160mm]{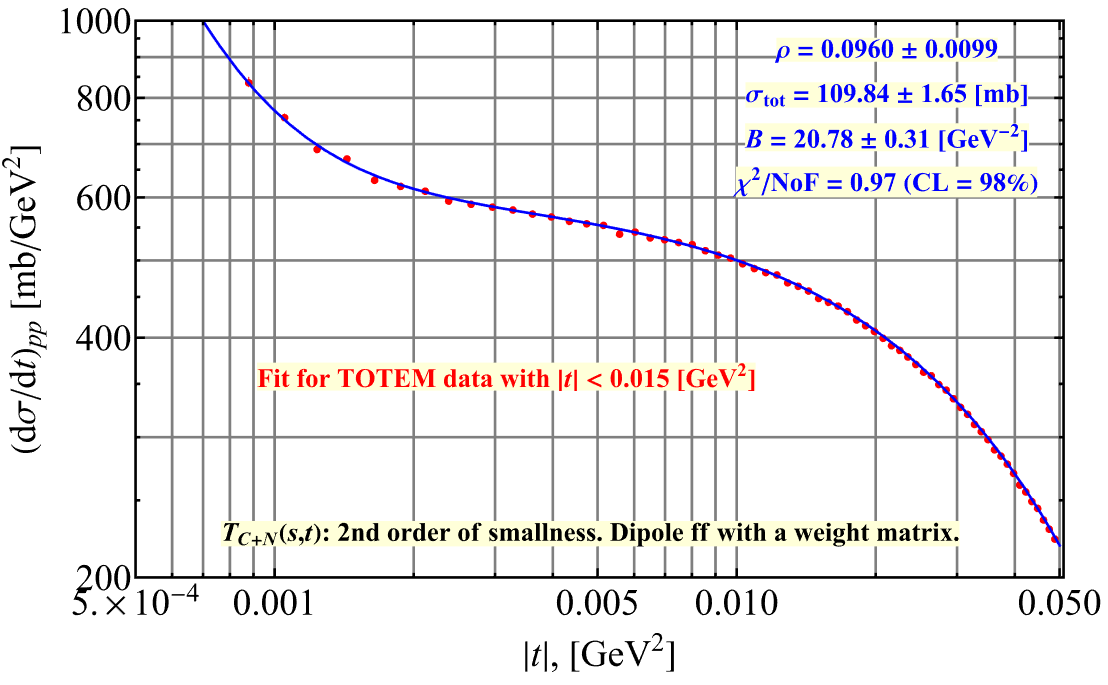}$$
\vspace{-8.6mm}
\caption{{\it Results of fitting the experimental data at}
$\sqrt{s}= 13$ TeV
{\it with} $|t| < t_0 \cong 0.015\mbox{ GeV}^{2}$. {\it Only statistical
errors are indicated.}
}
\label{Pic4}
\end{figure}

Thus, the values of the parameters, taking into account errors,
do not contradict their same values for small and large cut-off values. 
In the interval $ 0.01 < t_0 < 0.05 $, we observe the stability of the
parameters extracted from the fitting, and for this reason we use the
same method for choosing a particular fit as in the previous case,
i.e. we take a fit at $t_0\cong 0.015~  \mbox{GeV}^{2}$. The results of
this fit are shown in Fig. \ref{Pic4}.

 Extracted parameters are as follows: 

$$\rho= 0.10\pm 0.01,$$
$$\sigma_{\mbox{tot}}= 109.5 \pm 1.6~ [\mbox{mb}],$$
$$B = 21.02 \pm 0.26~ [\mbox{GeV}^{-2}].$$
We observe a good course of the curve along the experimental points, but
already shifted, each by $\lambda = -0.64106$ from its systematic error
(downward shift).

\newpage

\vspace{-6.1mm}
\subsection{The $\chi^{2}$-criterion with use of the "method of the shifting
experimental data" (variant 2).\vspace{-0.1mm}}
 
Let us consider one more way of shifting the experimental data. We will shift
the central values by a value proportional to the (measured) value itself,
and the proportionality coefficient $ \lambda $ is assumed to be the same
for all points. In this case $ \chi^{2} $  has in the form  
$$
\chi^{2} =  \sum_{i=1}^{N} \left[\frac{Y^{th}(x_{i})-\lambda Y^{exp}(x_{i})}
{{\Delta}Y_{stat}(x_{i})}\right]^{2} + \left(\frac{\lambda -1}{0.055}
\right)^{2},
$$
and $ \lambda $ is the same for all points of the array of experimental data.
The second term on the right is
a penalty function for $\chi^{2}$, which arises in statistics as a result
of an experimental data shift in systematic errors. This form is due to the
shifting method and the fact that all systematic errors are equal to
$0.055$ of the differential cross section.
That is, we shift  the experimental values up or down by an amount proportional
to the measured value itself.
It is clear that in this case $\lambda$ should be positive and close to unity.
We assume this factor to be the same for all experimental points. At the
same time, in the expression for $ \chi^{2} $, we use only the statistical
error of the measured value. The multiplier $ \lambda $ itself is a fitting
parameter.

Fitting in this case should not differ significantly from fitting when using
the weight matrix for $ \chi^{2} $ and its results are shown in
Fig. \ref{Pic5} (according to the formulas in Eq.(\ref{eq7}) -- (\ref{eq8})).

Here we have only three points with p-value $> 0.9.$ So we have no
choice but to choose a fit corresponding to the right extreme point
with a cross. As before, we observe in the interval
$0.01~ \mbox{GeV}^{2} < t_0 < 0.05~ \mbox{GeV}^{2}$,  the stability of
the parameters extracted from the fits, and for this reason we use the
same method of choosing a specific fit (although in this case we
actually have no choice). The results are shown in Fig. \ref{Pic6}.
The extracted parameters from this fit are as follows:
\newpage
\begin{wrapfigure}[24]{l}{104mm}
\noindent\includegraphics[width=104mm]{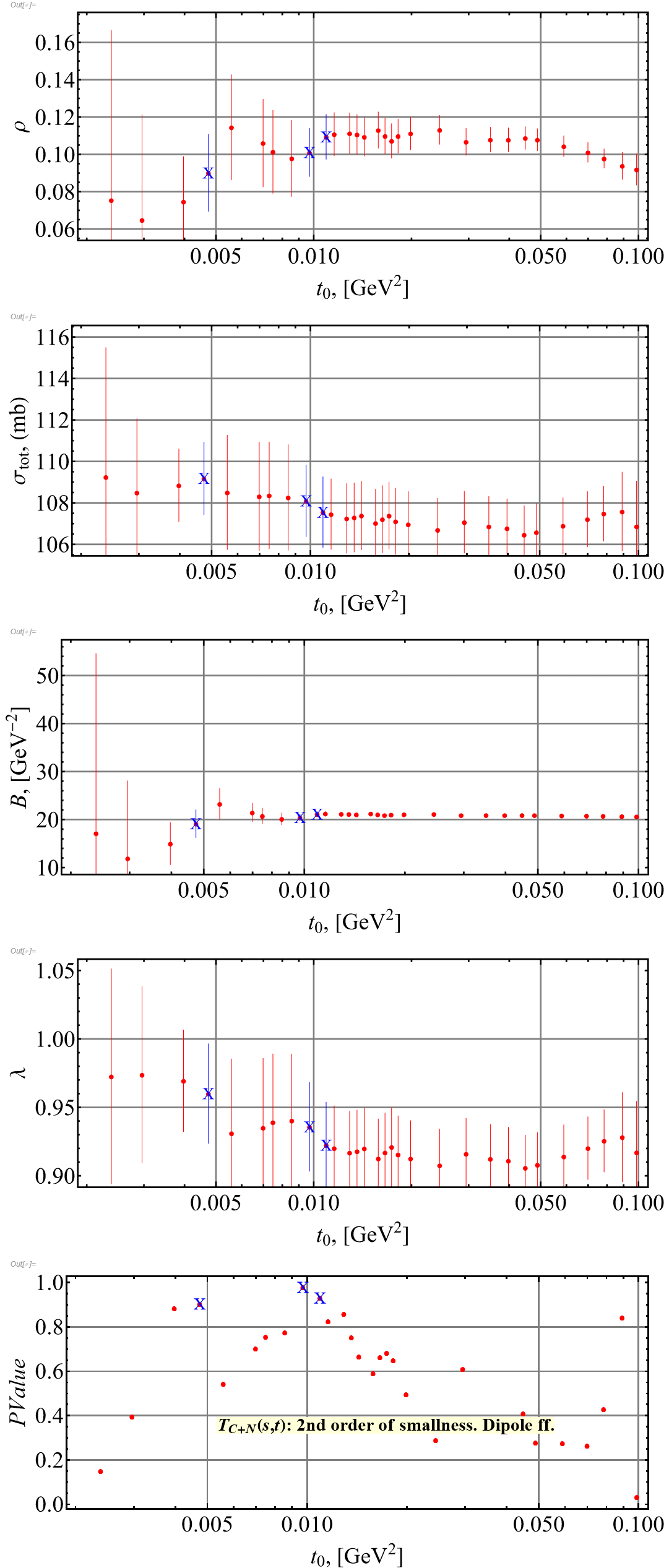}
\vspace{-8.6mm}
\caption{\textit{Parameter values for different cut-off  values $t_0$.
Points with p-value $ > 0.9 $ are marked with crosses.}}
\label{Pic5}
\end{wrapfigure}
\[\rho = 0.11 \pm 0.01,\vspace{-2.1mm}\]
\[\sigma_{\mbox{tot}} = 107.6 \pm 1.7\: [\mbox{mb}],\vspace{-0.1mm}\]
\[B = 21.15 \pm 0.55~[\mbox{GeV}^{-2}].\vspace{-0.1mm}\]

One can see a good course of the curve along the experimental points,
but already occupying a new central value equal to the coefficient
$\lambda = 0.920005$ times its old value (central values, as in the
previous case, are still moving down).

Thus, we have extracted the values of the parameters in three ways, and all
these three sets do not contradict each other (taking into account errors).

However, the values of the $\rho$-parameter do not allow making an
unambiguous statement that it is less than 0.1 as in
Ref. \cite{TOT} at $t_{0}= 0.07~\mbox{GeV}^{2}$.

\newpage

\begin{figure}[h]
\vspace{-4.1mm}
$$\includegraphics[width=140mm]{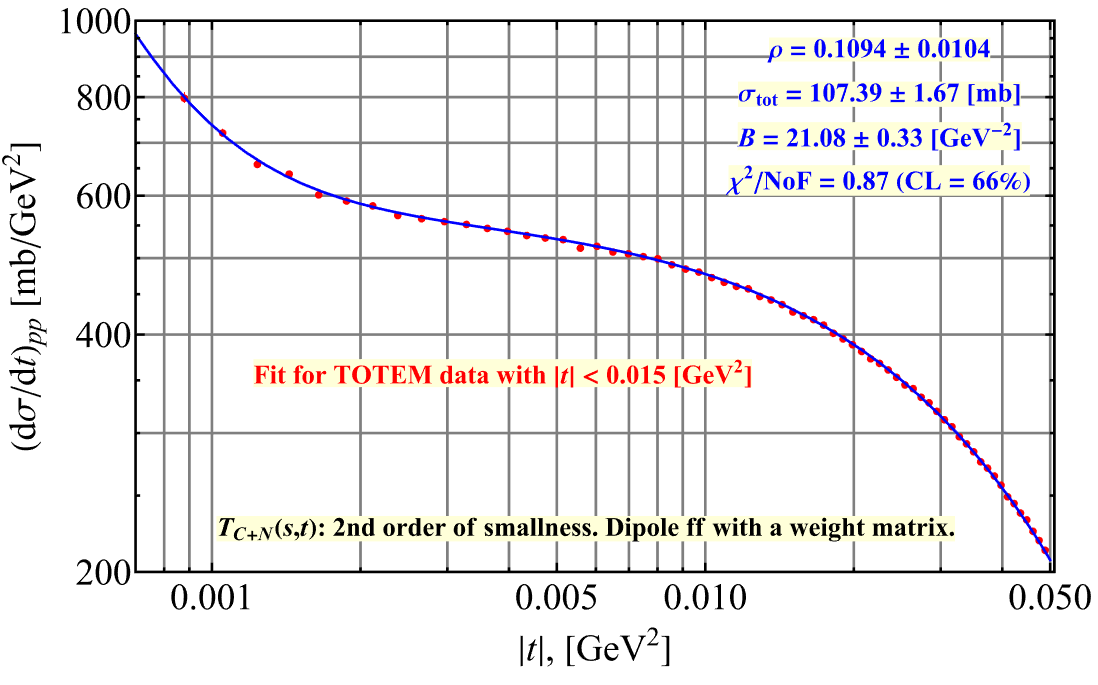}$$
\vspace{-11.2mm}
\caption{{\it Results of the fit of the experimental data at} $\sqrt{s} =
13\mbox{ TeV}$ {\it and}  $|t| \lesssim t_0 = 0.015~\mbox{ GeV}^{2}$.
{\it Only statistical errors are indicated.}}
\label{Pic6}
\end{figure}

\vspace{-9.9mm}
\subsection{The last step: cut-off of the experimental data with a small
value of $|t|$.\vspace{-2.1mm}}

In all the above cases of fitting by the method of cutting off points with
a high value of $ \vert t\vert $, there is a space where the extracted parameters
practically do not change. This happens in the interval
$0.01 \leq |t| \leq 0.05~\mbox{ GeV}^{2}$. However, at
$t_{0} < 0.01\mbox{ GeV}^{2}$  the values of the extracted parameters experience
sharp irregular changes. This is what motivates us to
discard the region $|t| < 0.01\mbox{ GeV}^{2}$. Of course, the data at
$|t| > 0.05\mbox{ GeV}^{2}$ should also be discarded, since above this
limit the extracted parameters begin to experience noticeable changes
and the confidence level becomes unacceptably low. 

\begin{figure}[h]
\vspace{-3.6mm}
$$\includegraphics[width=140mm]{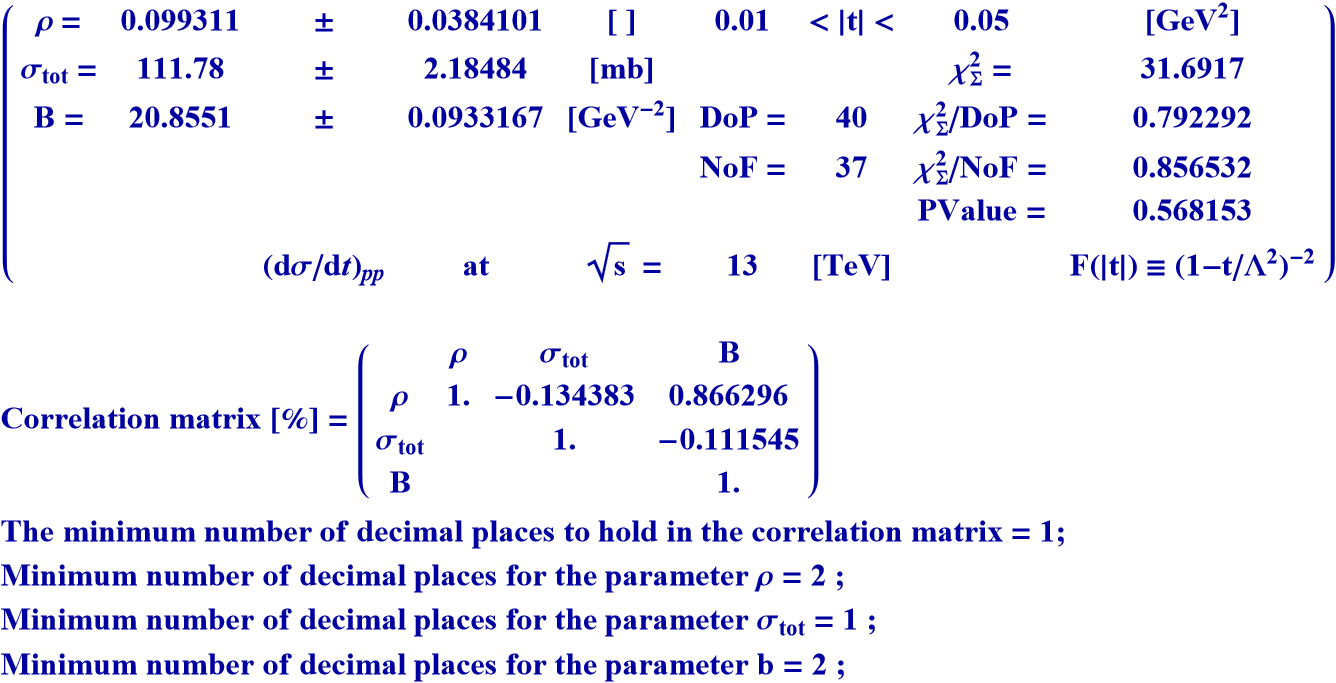}$$
\vspace{-9.1mm}
\caption{\textit{The summary of parameters.The number of decimal places to
retain in parameter values is given according to the arguments in}
\cite{Ezhela}. } 
\label{Pic7}
\end{figure}

\begin{figure}[h]
$$\includegraphics[width=165mm]{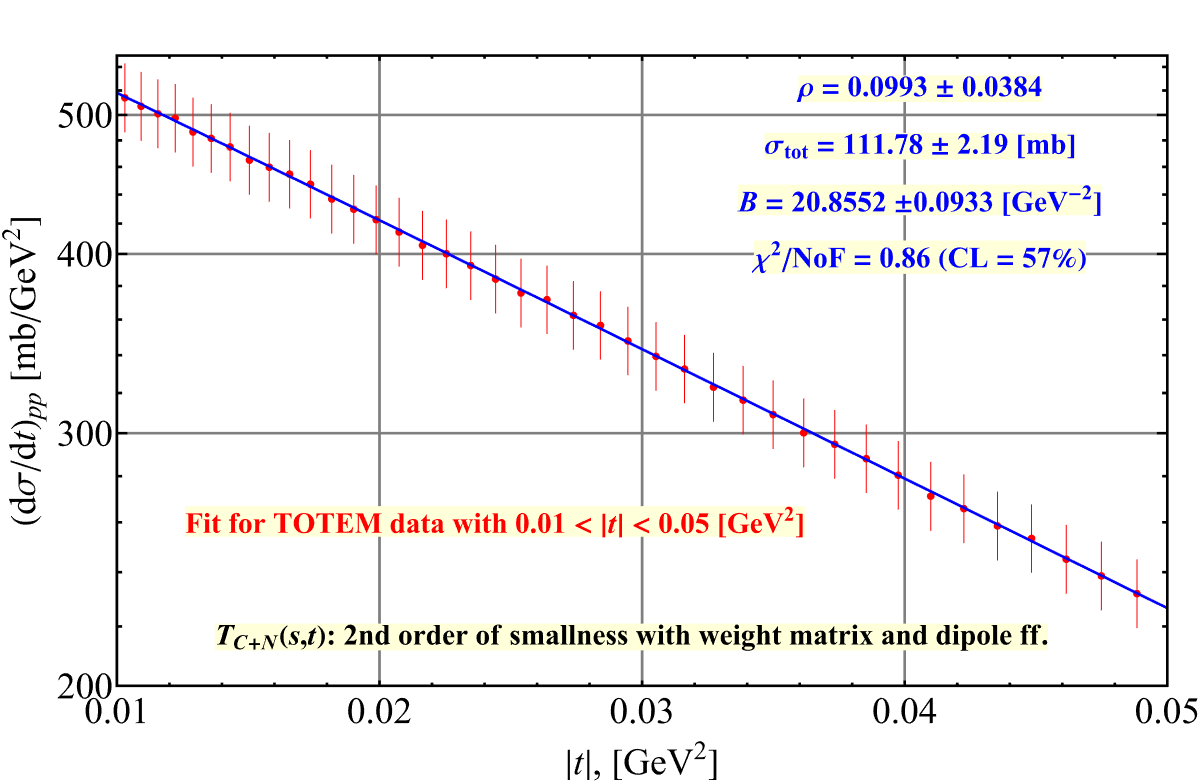}$$
\caption{{\it Results of experimental data fitting on the interval}
 $0.01 \leq |t| \leq 0.05\mbox{ GeV}^{2}$ {\it with
 use of the weight matrix. The scale along the vertical axis is
 logarithmic. The total errors extracted from the weight matrix are indicated.
 It can be seen that in this case there is no regular shift of the theoretical
 curve relative to the central experimental points, in contrast to the fitting
 on the full array (with a weight matrix).}}
\label{Pic8}
\end{figure}

Thus, we use only experimental data for which the stability of the extracted
parameters was observed above, i.e. the interval\vspace{-1.6mm}
$$
0.01\mbox{ GeV}^{2} \lesssim |t| \lesssim 0.05\mbox{ GeV}^{2}.
$$
The results of the fit of these data using the weight matrix are shown
in Fig. \ref{Pic7}.
\newpage
In Fig. \ref{Pic8} it can be seen that the theoretical curve and the
experimental data are in perfect agreement. Apparently, this result is
the most reliable on the TOTEM experimental data array at
$\sqrt{s}= 13$ TeV.

\begin{figure}[h]
$$\includegraphics[width=165mm]{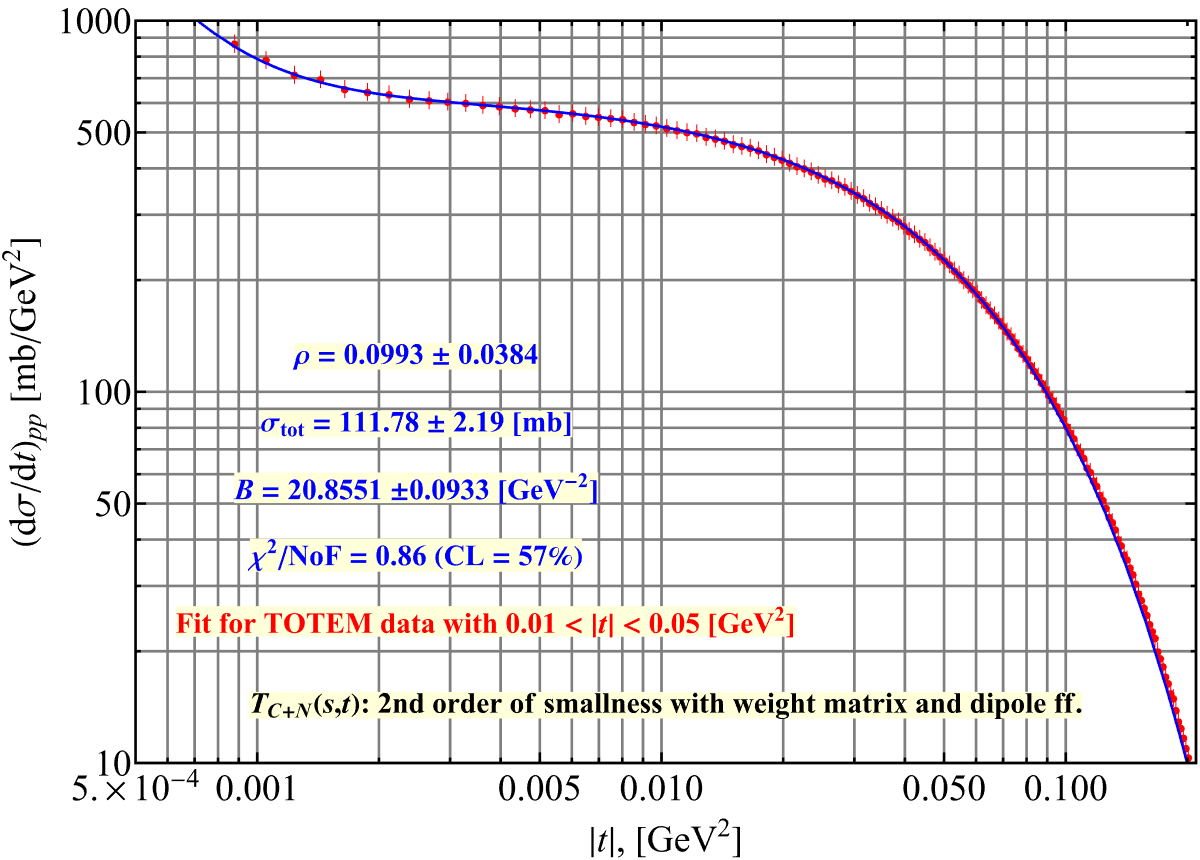}$$
\caption{{\it  Theoretical description of the full} TOTEM
{\it data on} $(d\sigma/dt)_{pp}$  {\it at} $\sqrt{s}=13$ TeV.
\textit{ The fit was carried out on the interval}
$0.01 \lesssim |t| \lesssim 0.05\mbox{ GeV}^{2}$.}
\label{Pic9}
\end{figure}

 For completeness, we present in Fig. \ref{Pic9}  the behaviour of the
descriptive curve on the entire array of experimental data\footnote{The reliability
of the cross-section description using the truncated series in $\alpha$
when lowering $ \mid t \mid $ is discussed in Appendix C.}.

The above fittings in two different ways with a shift in experimental data
give excellent statistical results and both indicate the possible fact of
overestimated central values of the experimental data of differential cross
sections. This is consistent with the first fitting result we cited above,
where the theoretical curve systematically passes below the central values.

Such a picture is nothing but a manifestation of the properties of the weight
matrix given by the TOTEM group. It is possible
to get rid of this contradictory
picture of the need to correct the experimental data (which, in our opinion,
is an unacceptable operation) only by limiting the TOTEM experimental data
sample to the segment $0.01\lesssim |t| \lesssim 0.05 \mbox{ GeV}^{2}$.
This leads to a new, different from the TOTEM, value of the
$\rho$-parameter and, most importantly, to its larger error ($\approx 40\%$),
which makes the results obtained in this way quite compatible with the results
of the COMPETE collaboration.

Thus, we carried out three types of processing of experimental TOTEM data,
described above by the method of a step-by-step exclusion of experimental
data with high values of $ q^{2} $. In all these methods, approximately the
same parameter values were obtained (see above), roughly near the values
of the TOTEM group.
A common property of these three methods is the fact that they exhibit a
high degree of parameter stabilization when using experimental data for which
$q^{2}\leq 0.05 GeV^{2}$, and with a high degree of confidence - more than
$85\%$. However, all three of these methods have significant drawbacks. In
the first method, the theoretical curve systematically passes below the experimental
values, while in the second and third methods, it is generally necessary
to shift the experimental values, and these methods, although they perfectly
describe the shifted experimental values, do not use the correlation (weight)
matrix given by the experimenters.

The most significant drawback of these three methods is the fact that when
discarding experimental points with a high $ q^{2} $ - value
($q^{2} \gtrsim 0.05\mbox{ GeV}^{2}$),
there is, in the area of
$q^{2}\lesssim 0.01\mbox{ GeV}^{2}$,  a significant decrease in the
confidence level of the obtained parameter values and their significant change
in [central] values (especially the parameter $\rho$) with a simultaneous
sharp increase retrieved parameter errors.

This prompted us, when processing the experimental values, to discard not
only the experimental data for which $ q^{2}\leq 0.05 GeV^{2} $, but also
those for which $q^{2}\leq 0.01 GeV^{2}$. Of course, we used the correlation
matrix (weight matrix) in this case. In this case, the results are obtained
with a high level of confidence and without the shortcomings that we mentioned
above. They are shown in Fig. \ref{Pic9}.

However, the price of getting rid of all the shortcomings of extracting
experimental data was a high error for the value of the $\rho$ - parameter,
which reaches about $40\%$. However, the theoretical curve does not systematically
pass above or below the experimental points (see Figures 8 and 9), and the
confidence level turns out to be even higher than $50\%$ .

\textit{For this reason, we consider these parameter values to be the most
consistent with the experimental measurements of the TOTEM group.}

Separately, we investigated the influence of the dipole form factor, which
we used everywhere above, on the results obtained. To do this, we carried
out similar calculations using the exact formulas for the unit form factor
("electrically point like protons"). The results turned out to be practically
indistinguishable from the above values of the parameters for the dipole
form factor. From this we conclude that the use of the
$\mathcal{O}(\alpha^{2})$-approximation for the description of the differential
cross-section is quite sufficient  at the range of the transferred momenta
considered.
\section*{Conclusions and outlook}
 In the present paper, a comprehensive discussion of various methods for
extracting the $ \rho $ parameter from the experimental results of the TOTEM
collaboration at $ \sqrt{s} = 13 $ TeV has been carried out. 
 
Theoretically substantiated formulas, Eqs.(\ref{eq4}) -- (\ref{eq5}) and,
for a certain general form for $T_{N}$ (Eq. \ref{eq6}),
Eqs. (\ref{eq7}) -- (\ref{eq8}),
are derived which allow us to describe the data for small values of
$q^{2}$ ($|t|$) .

It is shown that in order to extract the $\rho$ parameter from the experimental
data \cite{TOT} in a statistically sound manner, one can account
only for points that satisfy the condition
$0.01 \lesssim |t| \lesssim 0.05~\mbox{GeV}^{2}$.
Consequently, although the value of $ \rho $ remains to be close to 0.1,
its root-mean-square error increases, in contrast with the TOTEM result (10\%),
to almost 40\%.

For this reason, we consider it premature to conclude, as the authors of
many papers did, that the value of the  $\rho$ parameter was determined
by the TOTEM collaboration at  $\sqrt{s} = 13$ TeV  very accurately,
with $\rho = 0.09 \pm 0.01$ (from the array with
$|t| \leq 0.07~\mbox{GeV}^{2}$). 

It is also premature to draw far-reaching physical conclusions based on these
measurements, such as the existence of an Odderon or, at least,  its significant
contribution to the forward observables.

\textit{The aforesaid gives us also a reason to consider the conclusion about
the failure
of the COMPETE predictions \cite{PRED} announced in \cite{TOT} to be unfounded.}

Below we give the summary of the concrete numerical values of the basic
parameters obtained in this paper.
 
\subsubsection*{I}
  With use of the IR regular expression (\ref{eq7}) -- (\ref{eq8})
 for the differential cross-section $ d\sigma^{~pp}_{C+N}/dt $ which accounts
 in a correct way
 (in the $\mathcal{O} (\alpha^{2})$ approximation)  for the Coulombic contribution
 we have retrieved from the TOTEM data \cite{TOT} the following values of
 three important parameters at $ \sqrt{s}= 13$ TeV:
\begin{equation}
\sigma_{\mbox{tot}}^{~pp} = 111.8 \pm 2.2\mbox{ mb},
\label{eq9}
\end{equation}
$$
B^{~pp} = 20.86\pm 0.09\mbox{ GeV}^{-2},
$$
$$
\rho = 0.10\pm 0.04,
$$
$$
0.01 \lesssim |t| \lesssim 0.05\mbox{ GeV}^{2}.
$$
\subsubsection*{II}
In order to check the influence of the smearing of the electric charge over
the proton volume we have considered the idealized case of "electrically
point-like protons" ($\mathcal{F}=1$)and have retrieved the following
result:
\begin{equation}
\sigma_{\mbox{tot}}^{~pp} = 111.8 \pm 2.2\mbox{ mb},
\label{eq10}
\end{equation}
$$
B^{~pp} = 20.86\pm 0.09\mbox{ GeV}^{-2},
$$
$$
\rho = 0.10\pm 0.04,
$$
$$
0.01 \lesssim |t| \lesssim 0.05\mbox{ GeV}^{2}.
$$
It practically coincides with
Eq. (\ref{eq9})\footnote{Eqs. (\ref{eq9})
and (\ref{eq10}) actually differ in higher decimals but this, certainly,
does not matter much.}.
What could mean a coincidence of two sets (\ref{eq9}) and (\ref{eq10})?

The spread of the electric charge inside the proton is limited by the
physical size of the proton (of its valence core responsible for the form
factor at small $t$) which amounts to (in the transverse projection) about
0.6 fm \cite{OK} . The transverse interaction region is $(2B)^{1/2}$. At
$\sqrt{s}=13$ TeV the average size of the interaction region is about
1.3 fm . According to such a reasoning the proper size of the proton can
be hardly neglected in comparison with the average interaction radius.
So the mentioned coincidence seems a little puzzle. Although one could argue
that the form factors are essentially close to $1$ at considered $t$.
However, the form factors enter also the integrals in Eq.(5).\vspace{-2.1mm}
\subsubsection*{III}
Now we are to compare our results (9) with the published values
\cite{TOT}  retrieved on the basis of a different theoretical requisite
\cite{Cahn}\footnote{How this approximation differs from the exact approach
is shown in Appendix A. }  to treat CNI\vspace{-2.1mm}\footnote{From the
two sets of parameters given in \cite{TOT} we keep only the one that corresponds
to the upper cut-off $t_{0} = 0.07~\mbox{GeV}^{2}$ because stabilization
discussed in Section 4 takes place for $t_{0} < 0.1~\mbox{GeV}^{2}$.}: 
\begin{equation}
\sigma_{\mbox{tot}}^{pp} = 110.5 \pm 2.4\mbox{ mb},
\label{eq11}
\end{equation}
$$
B^{pp} = 21.78\pm 0.06\mbox{ GeV}^{-2},
$$
$$
\rho = 0.09\pm 0.01, 
$$
$$
|t| \leq 0.07\mbox{ GeV}^{2}.
$$
Note that the parameters $ \sigma_{\mbox{tot}}^{pp} $ and $b^{pp}$ of both
sets differ little ( $\leq 1\%$ )both in central values and in errors.
However, the values of $\rho^{pp}$ differ quite a lot. The difference is
$10\%$ for the central value and $30\%$ for the error band.

At first sight, our result Eq. (\ref{eq9}) looks quite ordinarily being,
in terms of
$\rho$,  close to the TOTEM result Eq. (\ref{eq11}) (although such a
difference may
have a significant physical reason). However, a conceptually important point
is that such a widening of the error corridor in Eq.(\ref{eq9}) in
compare with
Eq. (\ref{eq11}) "returns to the agenda" a number of models (giving 
$0.12\leq \rho \leq 0.14$, see e.g.\cite{Sel})which are to be rejected according
to \cite{TOT} .

Let us remind that in this paper we used in capacity of the nuclear amplitude
$T_{N}$ the option (\ref{eq6}) which, though widely used in literature (including
the TOTEM publication \cite{TOT}) , is, strictly speaking, not fully
satisfactory\footnote{This does not concern Eqs. (\ref{eq4}) -- (\ref{eq5})
which are of a general
character.} because of  $t$ -independence of its phase\footnote{The
importance of the $t$-dependence of the phase of the strong interaction amplitude
was addressed in Refs.\cite{Proch}, \cite{PePh}.}.  

We have also to notice that the values of $ \rho $ retrieved with help of
different models of $ T_{N} $ are inevitably different. Such a semi-theoretical
character of the $ \rho $-parameter was realized for a long time but never
studied systematically.

We hope to further investigate
these important issues in forthcoming studies.\vspace{-6.1mm}

\section*{Appendix A. On the difference between the $\alpha$-expansions
of the exact expression for $d\sigma_{C+N}/dt$ and the CKL
approximation.\vspace{-2.1mm}}

As was already mentioned above, the CKL scheme shows two flaws.

To begin with,
the account of the form factor in the formula for the full amplitude $T_{C+N}$
was assumed in a simplified way, different from the exact account.
Moreover, when using the expansion in $ \alpha $ CKL \cite{Cahn} retain in
the $T_{C+N}$ only the first order term. Since this approximation is then
used in $\mid T_{C+N}\mid^{2} \sim d\sigma_{C+N}/dt$ one must either use
only the $\mathcal{O}(\alpha)$-approximation also in $d\sigma_{C+N}/dt$
(which would eventually lead to an absurd situation) or, to be mathematically
correct, keep the $\mathcal{O}(\alpha^{2})$-term already in the amplitude.
Otherwise, there is a missing $\mathcal{O}(\alpha^{2})$ - term in
$d\sigma_{C+N}/dt$. 

It is not difficult to retain the $\mathcal{O}(\alpha^{2})$ - term in the
$\alpha$ -expansion of $T_{C+N}^{CKL}$ (ignoring the aforesaid blunder with
account of the form factor)and this is what we have done in order to make
the comparison between the two schemes better reflecting the essence of the
matter. Thus modified CKL cross-section (i.e. corrected with retaining of
the $\alpha^{2}$ term in the full amplitude) differs from the
$\alpha$ -expansion of the exact expression (\ref{eq4}), (\ref{eq5}) by
a decrement
$$
\delta \left[ \frac{d\sigma_{C+N}}{dt}\right]
\equiv \frac{d\sigma_{C+N}}{dt}-\frac{d\sigma^{CKL}_{C+N}}{dt}
$$
which takes place in the second order in $ \alpha$ only and reads as
follows
$$
\delta \left[ \frac{d\sigma_{C+N}}{dt}\right] = (\hbar c)^{2}
\frac{\alpha^{2}}{2\pi s q^{2}}\int
\frac{dk^{2}}{k^{2}}\frac{dp^{2}}{p^{2}}
(-\lambda (q^{2}, k^{2}, p^{2}))_{+}^{-1/2}\cdot
~~~~~~~~~~~~~~~~~~~~~~~~~~~~~~~~~~~~~~~~~~~~~~~~~~~~~~~~~~~~~~~~~~
$$
$$
~~~~~~~~~~~~~~~~~~~~~~
\cdot\left\{ q^{2}\mbox{Im} T_{N}(q^{2})\left[ \mathcal{F}^{2}(k^{2})\mathcal{F}^{2}(p^{2})-
\mathcal{F}^{2}(q^{2})\right] - k^{2} \mbox{Im} T_{N}(k^{2})\left[ \mathcal{F}^{2}(p^{2})
\mathcal{F}^{2}(q^{2})-\mathcal{F}^{2}(k^{2})\right] -\right.
$$
$$
~~~~~~~~~~~~~~~~~~~~~~~~~~~~~~~~~~~~~~~~~~~~~~~~~~~~~~~~~~~~~~~~~~~~~~~~~~~~~
\left. - p^{2}\mbox{Im} T_{N}(p^{2})\left[\mathcal{F}^{2}(q^{2})\mathcal{F}^{2}(k^{2})-
\mathcal{F}^{2}(p^{2})\right]\right\} +
$$
$$
~~~~~~~~~~~~~~~~~+ 
\frac{\alpha^{2}}{4\pi (\hbar c)^{2} s^{2} (2\pi)^{2}}\int\frac{d^{2}k}{k^{2}}
\frac{d^{2}p}{p^{2}}\left[\mathcal{F}^{2}\left((k+p)^{2}\right) - \mathcal{F}^{2}(k^{2})
\mathcal{F}^{2}(p^{2})\right]\cdot \mbox{Re}
\left[ T_{N}(q^{2})T^{*}_{N}\left( (q-k-p)^{2}\right)\right]
$$

In the point like limit $\mathcal{F} \rightarrow 1$,  the above integrals
turn out to be equal to zero and the CKL approximation (adjusted for the
$\alpha^{2}$-term retention in the amplitude) is equivalent to our formulas.

The decrement would also be equal to zero if the amplitude in
$T_{C+N}^{CKL}$ differed from $ T_{C+N} $ only by a phase factor\footnote{Under
the full amplitudes $ T_{C+N} $ we mean here the amplitudes after factoring
out the IR divergent part of the phase of the regularized amplitude with
a fictitious "photon mass"}. In this case, the amplitudes would be physically
equivalent. However, as we see, this is not the case. 

In general case the decrement behaves at small $q^{2}$ as
$$
\delta \left[ \frac{d\sigma_{C+N}}{dt}\right]\approx - \frac{\alpha^{2}}{16\pi}
\left\{ \left[\ln\left( \frac{\Lambda^{2}}{q^{2}}\right) -\frac{11}{6} -
A\left( \frac{B\Lambda^{2}}{2} \right)
\right]^{2}
\frac{\sigma^{2}_{\mbox{tot}}}{(\hbar c)^{2}}+ 64\pi
\mathcal{F}^{2}(q^{2})\frac{\sigma_{\mbox{tot}}}{\Lambda^{2}}
\right\}
$$
where
$$
A(z) = \int_{0}^{\infty}\frac{(1-e^{-z x})}{x (1+x)^{4}} dx =
\mathcal{C}+\frac{z^{2}-4z+\mbox{Ei}(-z)e^{z}(z^{3}-3z^{2}+6z-6)}{6}+
\ln z .
$$

In his talk at the Low-x 2021 Workshop at Elba \cite{Oest} K.\"{O}sterberg
had presented the comparison of our approach (as explained most recently
in \cite{PSt}) with the CKL approximate scheme \cite{Cahn}( adopted in \cite{TOT})
with a severe sentence(basing on the paper \cite{Kasp}):
\begin{center}
"the  \textit{new CNI formula from
Petrov... fails}".\vspace{-1.1mm}
\end{center}

Actually the comparison was made not in terms of the ratio
$d\sigma_{C+N}/d\sigma_{C+N}^{CKL}$ but, instead,  of the ratio
$d\sigma_{C+N}^{1}/d\sigma_{C+N}^{CKL}$ where
 $d\sigma_{C+N}^{1}/dt$  is the cross-section with \textit{only the first-order
 contribution in $ \alpha $} in the full amplitude and which was called the
 "new CNI formula from Petrov" in spite of that we specially emphasized the
 role of the second order in \cite{PSt}.

The ratio $ d\sigma_{C+N}^{1}/d\sigma_{C+N}^{CKL} $ showed a ($ 4\% $) deviation
from unity (in the region of low t) which was misunderstandingly qualified
as a failure of our approach \cite{PTR}, \cite{PSt}.

\begin{figure}[h]
$$\includegraphics[width=160mm]{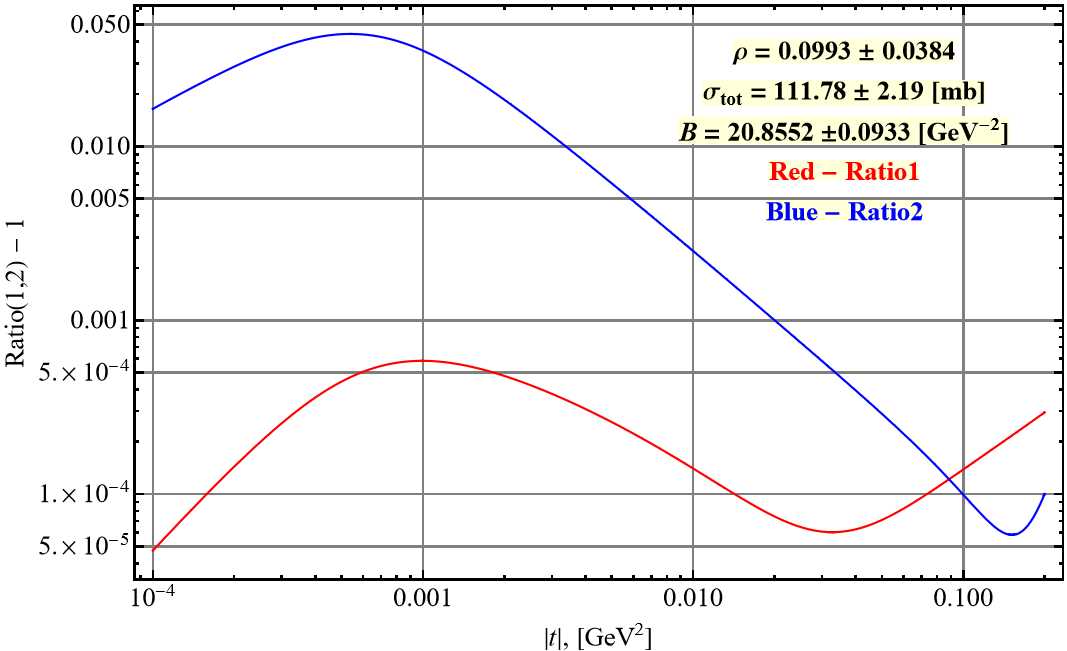}$$
\caption{\textit{The comparative quality of the expressions
$d\sigma_{C+N}^{1}/dt$ and $d\sigma_{C+N}^{CKL}/dt$ (which retain only
the first order terms in $ \alpha $ in the amplitude $T_{C+N}$) w.r.t.
the reference cross-section $ d\sigma_{C+N}/dt$ (Eq. (\ref{eq7})) with
a full account for $\alpha^{2}$ terms. }}
\label{Pic10}
\end{figure}

To clear things up let us, in our turn, estimate the comparative quality
of the quantity $d\sigma_{C+N}^{1}/dt$ and the TOTEM working tool, the CKL
expression $ d\sigma_{C+N}^{CKL}/dt $ \cite{TOT}, \cite{Cahn}.
At Fig. \ref{Pic10} we show (similar to what was done in \cite{Kasp}) the
ratios $ (d\sigma/dt - ref)/ref $ where we assume as a reference value,
$ref$, our expression in Eq. (\ref{eq7}) while for $d\sigma/dt$ we take
$d\sigma_{C+N}^{CKL}/dt$ and $d\sigma_{C+N}^{1}/dt$.
\footnote{with $d\sigma_{C+N}/dt$ from Eq.(\ref{eq7}) and
$$
 L(q^{2}) = \ln \left( \frac{\Lambda^{2}}{q^{2}}+1\right) -\sum_{k=1}^{3}
k^{-1} \left( 1+\frac{q^{2}}{\Lambda^{2}}\right)^{-k}
$$
}:

\newpage

$$
\mbox{Ratio 1} = 
d\sigma_{C+N}^{CKL}/d\sigma_{C+N} - 1 =
~~~~~~~~~~~~~~~~~~~~~~~~~~~~~~~~~~~~~~~~~~~~~~~~
~~~~~~~~~~~~~~~~~~~~~~~~~~~~~~~~~~~~~~~~~~~~~~~~~
$$
\[ ~~~
= \alpha^{2}\left\{ \frac{\sigma^{2}_{\mbox{tot}}e^{-Bq^{2}}
\left[ A(B\lambda^{2}/2)-
\ln(\Lambda^{2}/q^{2})+11/6\right]^{2}}{16\pi (\hbar c)^{2}} +
\frac{4\mathcal{F}^{2}(q^{2})\sigma_{\mbox{tot}}\ln (1+q^{2}/\Lambda^{2})
e^{-Bq^{2}/2}}{q^{2}} \right\}/\left(\frac{d\sigma_{C+N}}{dt}\right)
\]
\[
\mbox{Ratio 2} = 
 d\sigma_{C+N}^{1}/d\sigma_{C+N} - 1 =
~~~~~~~~~~~~~~~~~~~~~~~~~~~~~~~~~~~~~~~~~~~~~~~~~
~~~~~~~~~~~~~~~~~~~~~~~~~~~~~~~~~~~~~~~~~~~~~~~
\]
\[
~~~~~~~~~~~~~~~~~~~~~~~=\alpha^{2} \left\{
\frac{\sigma_{\mbox{tot}}^{2}(1+\rho^{2})e^{-Bq^{2}}
\left[ A(B\Lambda^{2}/2)-D\right]^{2}}{16\pi(\hbar c)^{2} }
+ \frac{\mathcal{F}^{2}(q^{2})
\sigma_{\mbox{tot}}e^{-Bq^{2}/2}L(q^{2})}{q^{2}} \right\}
/\left(\frac{d\sigma_{C+N}}{dt}\right)
\]

The deviation from the reference values as seen at Fig. \ref{Pic10}
demonstrates that the account for the $\mathcal{O}(\alpha^{2})$ terms
\textit{in the amplitude} gets more and more essential with decreasing
$\mid t \mid$ . Hereof their influence on the forward observables,
first of all the $\rho$-parameter.
 
How to explain an evident imparity of the two ratios in Fig.10 with respect
to account for the $ \alpha^{2} $ terms? If two amplitudes differ in their
phases only they may differ significantly in the form of the amplitude. Nonetheless,
they give the same result  for the cross-section, both in all-order or when
retaining a few terms in $ \alpha $. However, if we deal with physically
non-equivalent amplitudes, the results may differ significantly. In particular,
non-equivalent amplitudes differently react to the inclusion of higher orders
in $\alpha$. What we deal with in our case is an evident physical non-equivalence
of the two amplitude moduli in question.

\section*{Appendix B. Point like electric charges.}
We believe it is instructive to consider the idealized case of "electrically
point like" protons with $\mathcal{F}=1$. In this case one can obtain explicit
all-order expressions.

In case of "electrically point like" protons, i.e. if $\mathcal{F} = 1$,
we would have a compact explicit expression for the Coulomb kernel
$S^{C} (q^{'},q^{''})$ expressed in terms of the well known generalized
functions described, e.g., in \cite{Gel}:
$$
S^{C} (q^{'},q^{''}) = 
(4\pi\alpha)^{2} \frac{(q^{''2}/q^{'2})^{i\alpha}}{q^{'2}q^{''2}}.
$$
Physically the use of such an approximation would be justified if the average
impact parameter between the colliding protons were much larger than the
"charge radius" of the proton. If (with some reservation) we take as the
average transverse distance between the centres of colliding protons the
value $ \sqrt{2B(s)}\approx \langle b^{2} \rangle^{1/2}$,  where $B(s)$
is the forward slope, then it exceeds (at $ \sqrt{s} = 13 $ TeV) the size
of the "valence core" ($\approx 0.63$ fm) \cite{OK}  only two times.

Nonetheless, we will analyse, in the end of this Section,  such an option
vs the TOTEM data in order  that afterwards, when using a realistic
form factor  ($ \mathcal{F} \neq 1 $), to see better the influence of
smearing of the electric charge of the proton. 

Eqs.(\ref{eq2}) and (\ref{eq3}) allow to transform Eq.(\ref{eq1}) into an
all-order (in $\alpha$) expression
$$
\frac{d\sigma_{C+N}}{dt} =
\frac{(1+\rho^{2})\sigma^{2}_{\mbox{tot}}\mbox{exp}(-Bq^{2})}
{16\pi (\hbar c)^{2}}\cdot
\left| ~\Gamma (1+i\alpha)~
_{1}F_{1}(i\alpha,1;z)~\right|^{2} -
~~~~~~~~~~~~~~~~~~~~~~~~~~~~~~~~~~~~~~~~~~~~~~~~~~
$$
\begin{equation}
-\alpha\frac{\sigma_{\mbox{tot}}}{q^{2}}\mbox{exp}\left(\frac{-Bq^{2}}{2}\right)
 \cdot \mbox{Re}\left\{ (\rho-i)\cdot \exp\left[ -i\alpha
 \ln\left(\frac{Bq^{2}}{2}\right)\right]
\cdot\Gamma (1+i\alpha)~
_{1}F_{1}(i\alpha,1;z) \right\} +
\alpha^{2} \frac{4\pi (\hbar c)^{2}}{q^{4}}
\label{eq12}
\end{equation}
Here $_{1}F_{1}(i\alpha,1;z)$ is one of the confluent hyper geometric
functions\footnote{~\vspace{-4.1mm}
$$
_{1}F_{1}(i\alpha,1;z)=
\frac{1}{\Gamma(i\alpha)\Gamma (1-i\alpha)}\int_{0}^{1}
dx e^{zx} x^{i\alpha -1} (1-x)^{-i\alpha} .
$$
}
(see Chapter 9.21 in \cite{Grad}):
At energies up to 13 TeV and $q^{2}= -t\lesssim\ 0.05~\mbox{GeV}^{2}$
the values of $ z = B(s)q^{2}/2 $ if of order 0.5 or less. At such
$z$ the function $\Gamma (1+i\alpha)~_{1}F_{1}(i\alpha,1;z)$ is extremely
close to 1 and is exhaustively approximated by the following expression
$$
\Gamma (1+i\alpha) _{1}F_{1}(i\alpha, 1;z) = 1 +i\alpha f(z)
-\alpha^{2} g(z)
$$
where
\[f(z) = \sum_{n=1}^{\infty}\frac{z^{n}}{n n!}- \mathcal{C}\] and~~
\[g(z) = \sum_{n=1}^{\infty}
\frac{z^{n} \psi (n)}{n n!}  +\frac{\mathcal{C}^{2}}{2} +
\frac{\pi^{2}}{12} , ~~~\mathcal{C} = 0.5772...\]
With account of Eqs. (\ref{eq4}) and (\ref{eq5}) we get the following expression
for the differential cross-section
\begin{equation}
\frac{d\sigma_{C+N}}{dt} \left[ \mbox{mb/GeV}^{2} \right] =
\frac{(1+\rho^{2})\sigma_{\mbox{tot}}^{2}}{16\pi (\hbar c)^{2}}
e^{Bt} -\alpha \frac{\rho\cdot
\sigma_{\mbox{tot}}}{\mid t\mid}e^{Bt/2} +
~~~~~~~~~~~~~~~~~~~~~~~~~~~~~~~~~~~~~~~~~~~~~~~~~~~~~~~~~~~~~~
\label{eq13}
\end{equation}
\[
~~~~~~~~~~~~~~~~~~~~~~~~~~~~~~~~
+ \alpha^{2}\left\{ \frac{4\pi(\hbar c)^{2}}{t^{2}}
-\frac{\sigma_{\mbox{tot}}}{|t|}e^{Bt/2} \left[ f(z)- \mbox{ln}z \right]
+\frac{(1+\rho^{2})\sigma_{\mbox{tot}}^{2}}{16\pi(\hbar c)^{2} }
e^{Bt}\left[ f^{2}(z)-2g(z) \right]\right\}.
\]
We  remind that $z = B(s)q^{2}/2 \equiv Bq^{2}/2\equiv -Bt/2 \equiv B |t|/2.$

Let us use Eq.(\ref{eq12}) for description of the TOTEM data in the Coulomb-nuclear
interference region.
We omit repeating of several steps similar to those explained in the case
of realistic form factors and show the final result. We also do not include
the plots which are practically identical to those with the dipole form factor.

So, the values of the three standard parameters
resulting from the description of the data under the assumption of "electrically
point-like protons" ($\mathcal{F}=1$) are as follows:
$$
\sigma_{\mbox{tot}}^{~pp} = 111.8 \pm 2.2\mbox{ mb},
$$
$$
B^{~pp} = 20.85\pm 0.09\mbox{ GeV}^{-2},
$$
$$
\rho = 0.10\pm 0.04,
$$
$$
0.01 \lesssim |t| \lesssim 0.05\mbox{ GeV}^{2}.
$$

We have to emphasize that the use of the $\mathcal{O}(\alpha^{2})$
approximation appears practically equal to the use of the exect
expression (\ref{eq11}) (in more detail see Eq. (\ref{eq17})).

When looking at the values of parameters
$\rho,~\sigma_{\mbox{tot}},~B$
one should pay attention that in spite of our ignoring the form factor
effects the results almost coincide with those obtained  with full
account of the form factor.

In Appendix A we remarked that the CKL scheme (if corrected with inclusion
of the second order term in the amplitude) in the point like limit (at
$\mathcal{F}=1$) is identical to the ours, i.e. the correct one.
Here, however, we take for comparison the \textit{original} expression for
the amplitude in the point like case \cite{Cahn}. 

The 1st order CKL amplitude for a point like electric charge is of the
form (we use a special designation $ \hat{T}^{CKL}_{C+N} $ for an identical
remake of the corresponding expressions for the full amplitude in the original
versions \cite{Cahn}, viz Eq. (17) in the first reference in \cite{Cahn}
and Eq. (24) in the second one:
\[\hat{T}^{CKL}_{C+N}\mid_{\mathcal{F}=1} = T_{N }(q^{2}) -
\frac{8\pi \alpha s}{q^{2}} + i\alpha \int_{0}^{p^{2}}
\frac{dk^{2}}{k^{2}}[ T_{N}(q^{2}) - \bar{T}_{N}(k^{2}, q^{2})].
\]
So, for $T_{N}$ as in Eq. (\ref{eq6}) we have explicitly
\begin{equation}
\hat{T}^{CKL}_{C+N} = T_{N }(q^{2}) - \frac{8\pi \alpha s}{q^{2}}+
 i\alpha T_{N }(q^{2})\left[ \ln \left(\frac{Bp^{2}}{2}\right)+
 \mathcal{C} -
 \sum_{n=1}^{\infty}\left(\frac{Bq^{2}}{2}\right)^{n}/n! n\right] .
\label{eq14}
\end{equation}

As we can see, the limit $ s\rightarrow\infty $ in the upper  integration
limit is no longer possible and we get
$$
\frac{d\hat{\sigma}_{C+N}^{CKL}}{dt}\mid_{\mathcal{F}=1} =
\frac{\sigma^{2}_{\mbox{tot}} e^{-Bq^{2}}}{16\pi (\hbar c)^{2} } -
\alpha \frac{\rho (s)
\sigma_{\mbox{tot}}}{q^{2}}e^{Bq^{2}/2} +
\alpha^{2} \left\{ \frac{4\pi (\hbar c)^{2}}{q^{4}} \right.
+~~~~~~~~~~~~~~~~~~~~~~~~~~~~~~~~~~~~~~~~~~~~~~~~~~~~~~~~~~~~
$$
$$
~~~~~~~~~~~
\left.
+\frac{\ln \left(\frac{Bp^{2}}{2}\right)+\mathcal{C} -
\sum_{n=1}^{\infty}\left(\frac{Bq^{2}}{2}\right)^{n}/n! n}{q^{2}}
\sigma_{\mbox{tot}}e^{-Bq^{2}/2}+
\frac{\left[\ln \left(\frac{Bp^{2}}{2}\right)-
\sum_{n=1}^{\infty}\left(\frac{Bq^{2}}{2}\right)^{n}/n! n\right]^{2}}
{16\pi (\hbar c)^{2}}
\sigma^{2}_{\mbox{tot}}e^{-Bq^{2}}
\right\} ,
$$
where $\mathcal{C}=0.5772...$.

If we take equation for $d\hat{\sigma}_{C+N}^{CKL}/dt$ for
fitting the data we get the following result for our basic parameters
\begin{equation}
\sigma_{\mbox{tot}}^{~pp} = 111.81 \pm 1.75 \mbox{~mb},
\label{eq15}
\end{equation}
$$
B^{~pp} = 20.76\pm 0.30\mbox{ GeV}^{-2},
$$
$$
\rho = 0.2740\pm 0.0098,
$$
$$
\mid t\mid \lesssim 0.015 \mbox{ GeV}^{2}.
$$
In this case, the curve systematically passes below the central values of
the experimental points (the PPP effect, see \cite{ppp}), which, as before,
indicates that the systematic errors are overdetermined. However, the value
of the $\rho$-parameter turns out to be unexpectedly high. This probably
indicates an incomplete account for  terms of the second order in
$\alpha$. Nonetheless, the values of $\sigma_{\mbox{tot}}$ and $B$ turn
out to be quite consistent (within the error) with the previously obtained
values.
 
 Thus, we see that the point like limit according to \cite{Cahn} leads to
 the values of the basic parameters Eq. (\ref{eq15}) different from those
 in the case
 of the dipole form factor, cf with Eq. (\ref{eq11}). However, this mainly
 concerns  the $\rho$ parameter while the other two seem to be little sensitive
 to the change in the form factor account.
 
  As above, we also have made a fit over the most confidential interval of
experimental data values, i.e. at $0.01 \leq |t| \leq
0.05~\mbox{GeV}^{2}$. In this case, the curve passes ideally relative to
the experimental points, the parameters $\sigma_{\mbox{tot}}$ and $B$, as
elsewhere above, show enviable stability, while the $\rho$-parameter
undergoes a significant change remaining, however, unusually
high.
\begin{equation}
\sigma_{\mbox{tot}}^{~pp} = 112.36 \pm 2.18 \mbox{~mb},
\label{eq16}
\end{equation}
$$
B^{~pp} = 20.54\pm 0.06\mbox{ GeV}^{-2},
$$
$$
\rho = 0.2356 \pm 0.0332,
$$
$$
0.01 \lesssim |t| \lesssim 0.05 \mbox{ GeV}^{2}.
$$
Howbeit, we would like to point out again that the formula Eq. (\ref{eq14})
is the result
of an incorrect manipulation with divergent integrals and therefore contains
UV logarithms, which are unnatural for problems related to diffraction and
Coulomb-nuclear interference, as already indicated earlier
in \cite{PTR}.

\section*{Appendix C. Expansion in $ \alpha $: where to stop?}

Here we give some comments on the use of perturbative expansion
in the fine structure constant $\alpha$ in the narrow sense i.e. as applied
to the concrete problem considered in the text.

As to the general opinion about QED as a perturbation theory, there is a
conviction that the series in  $\alpha$ does not converge and should be considered
as an asymptotic one due to notorious non-analyticity in coupling constant
at the origin. This does not preclude to some parts of the amplitudes to
have convergent expansion in $ \alpha $ while non-analytic terms are of a
minor significance (like, say, exp$(-c/\alpha), c > 0$).
This is exactly what we have in Eq.(1) where the expansion in $\alpha$
has an infinite convergence radius and we tacitly assumed the absence of
non-analytic terms.

However, in practice, when one deals with a truncated series in
$\alpha$ one has to estimate the error of the series truncation used,
i.e. to estimate the value of discarded terms. It may well happen that for
some values of the additional parameters (in our case $s$ and $t$) the
approximation is justified, while for others it cannot be considered as an
"approximation" at all. In the latter case one, generally,  has to account
for more terms in expansion parameter.

As a simple example let us consider an oversimplified account of the point-like
Coulomb interaction when the full scattering amplitude is given as
follows
\[
T_{C+N}(s, t) = T_{N}(s, t) + \alpha \cdot 8\pi s/t.
\]
If (just for illustration) we take the parameters from the Conclusion and
neglect the $\rho$
value we obtain that the second "correction" term at the "critical value"
$|t|= 6.4\cdot 10^{-4}\mbox{ GeV}^{2}$ becomes equal to the "main"
and then, at lower $|t|$ surpasses it. Let us note that the lowest
$|t|$ reached by the TOTEM Collaboration (at $\sqrt{s} = 13$ TeV)
is $|t|= 8.79\cdot 10^{-4}\mbox{ GeV}^{2}$. This is already a signal
that it is necessary to take into account at least the second order in
$\alpha$. At the same time, at $\mid t \mid = 10^{-2}\mbox{ GeV}^{2}$
the use of the first order Coulomb term is fairly justified and constitutes
near $6\%$ of the main term $T_{N}$.

Sure, in more realistic  cases the critical values of $|t|$ may
be different but the conclusion  is the same: the weight of terms
in the $\alpha$ -expansion is  "running" with $|t|$.

\begin{figure}[h]
$$\includegraphics[width=150mm]{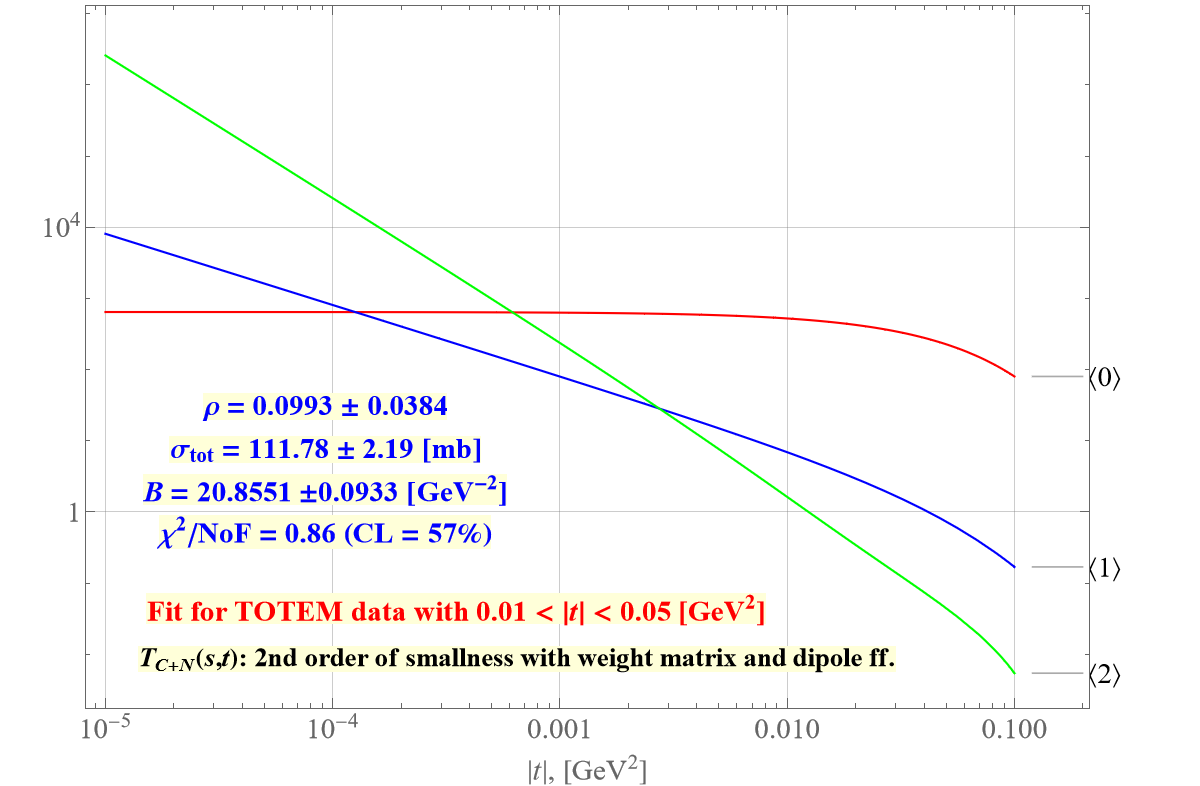}$$
\caption{Contributions of different orders of $\alpha$ in
$d\sigma_{C+N}/dt$.}
\label{Pic11}
\end{figure}

Figure 11 shows that for $|t| > 0.003\mbox{ GeV}^{2}$, we get the
"natural" order (we designate the n-th order as $\langle n \rangle$)
i.e.
$$
\langle 0\rangle > \langle 1\rangle > \langle 2\rangle.
$$
At $6\cdot 10^{-4}\mbox{ GeV}^{2} < |t| < 3\cdot10^{-3}\mbox{ GeV}^{2}$
the order breaks down:
$$
\langle 0\rangle > \langle 2 \rangle > \langle 1\rangle
$$
and, finally, at $|t| < 1.3\cdot10^{-4} \mbox{ GeV}^{2}$ the "natural"
order is broken even more, actually is reversed:
$$
\langle 2\rangle > \langle 1 \rangle > \langle 0\rangle.
$$
This circumstance prompts us to check the error determined by the remainder
term of the expansion in a series in $ \alpha $ which, in this case, is the
expression
\begin{equation}
\delta_{3} \frac{d\sigma_{C+N}}{dt} = - \alpha^{3}
\frac{\rho \sigma_{\mbox{tot}}e^{-Bq^{2}/2}} {q^{2}} \left[
\frac{3}{2} \ln^{2}\left(\frac{2}{Bq^{2}}\right) -
\ln\left(\frac{2}{Bq^{2}}\right)\left(\mathcal{C}+\frac{1}{2}-
Bq^{2}\right) - \frac{\pi^{2}}{12} \right].
\label{eq17}
\end{equation}

Using parameters from Eq.(9) we are convinced that the error does not exceed
a fraction of a percent of the full expression in the entire range of experimentally
available values of $t$.

Thus for
$10^{-4}\mbox{ GeV}^{2} \leq |t| \leq 5\cdot 10^{-2}\mbox{ GeV}^{2}$
higher orders in $ \alpha $ are quite harmless and the use of Eq.(7)
is fairly justified.

\section*{Acknowledgements}
We are grateful to  Vladimir Ezhela, Anatolii Likhoded, Jan Ka\v{s}par,
Vojtech Kundr\`{a}t  for their interest to this work and inspiring
conversations and correspondence.
We thanks V.Yu.Rubaev for a valuable help.

\end{document}